\documentclass[12pt,preprint]{aastex}

\newcommand\ugrizprime{u^{\prime}g^{\prime}r^{\prime}i^{\prime}z^{\prime}}
\newcommand\cfhtugriz{u^{*}g^{\prime}r^{\prime}i^{\prime}z^{\prime}}
\newcommand\ustar{u^{*}}
\newcommand\uprime{u^{\prime}}
\newcommand\gprime{g^{\prime}}
\newcommand\rprime{r^{\prime}}
\newcommand\iprime{i^{\prime}}
\newcommand\zprime{z^{\prime}}
\newcommand\Teff{T_{\rm{eff}}}

\newcommand\FeH{\rm{[Fe/H]}}

\received{}
\accepted{}
\journalid{}{}
\articleid{}{}

\slugcomment{Astronomical Journal, submitted}

\shortauthors{Clem et al.} 
\shorttitle{$\ugrizprime$ Fiducial Sequences}

\begin{document}

\title{FIDUCIAL STELLAR POPULATION SEQUENCES \\ FOR THE $\ugrizprime$ 
SYSTEM$^{1}$\footnotetext[1]{Based on observations obtained with MegaPrime/MegaCam, a joint 
project of CFHT and CEA/DAPNIA, at the Canada-France-Hawaii Telescope (CFHT) which is operated by 
the National Research Council (NRC) of Canada, the Institut National des Science de l'Univers of 
the Centre National de la Recherche Scientifique (CNRS) of France, and the University of 
Hawaii.}}

\author{James L.~Clem\altaffilmark{2} and Don A.~VandenBerg}
\affil{Department of Physics \& Astronomy, 
       University of Victoria, 
       P.O.~Box 3055, 
       Victoria, B.C. V8W~3P6, Canada}
\email{jclem@uvastro.phys.uvic.ca, vandenbe@uvic.ca}

\and

\author{Peter B.~Stetson}
\affil{Dominion Astrophysical Observatory, 
       Herzberg Institute of Astrophysics, 
       National Research Council, 
       5071 West Saanich Road, 
       Victoria, BC V9E 2E7, Canada}
\email{Peter.Stetson@nrc.gc.ca}

\altaffiltext{2}{Current address:  Department of Physics \& Astronomy, Louisiana State 
University, 202 Nicholson Hall, Baton Rouge, LA  70803 USA;  jclem@phys.lsu.edu}

\begin{abstract}

We describe an extensive observational project that has obtained high-quality and homogeneous 
photometry for a number of different Galactic star clusters (including M$\,$92, M$\,$13, M$\,$3, 
M$\,$71, and NGC$\,$6791) spanning a wide range in metallicity ($-2.3\lesssim\FeH\lesssim+0.4$), as 
observed in the $\ugrizprime$ passbands with the MegaCam wide-field imager on the 
Canada-France-Hawaii Telescope.  By employing these purest of stellar populations, fiducial 
sequences have been defined from color-magnitude diagrams that extend from the tip of the red-giant 
branch down to approximately 4 magnitudes below the turnoff: these sequences have been accurately 
calibrated to the standard $\ugrizprime$ system via a set of secondary photometric standards 
located within these same clusters.  Consequently, they can serve as a valuable set of empirical 
fiducials for the interpretation of stellar populations data in the $\ugrizprime$ system.

\end{abstract}

\keywords{Hertzprung-Russell diagram --- globular clusters: general --- globular clusters: 
individual (M$\,$92, M$\,$13, M$\,$3, M$\,$71) --- open clusters and associations: individual 
(NGC$\,$6791)}

\section{Introduction}

Recently, the Sloan Digital Sky Survey (SDSS) officially ended its planned five years of sky 
scanning operations to obtain an unprecedented amount of imaging and spectroscopic data for 
approximately one-quarter of the sky.  The SDSS was carried out on a dedicated 2.5$\,$m telescope 
equipped with a large-format mosaic CCD to image the entire northern Galactic cap (i.e., 
$b>30^{\circ}$) in five photometric bands and two digital spectrographs to provide spectra for 
$\sim1$ million stars, galaxies, and quasars scattered throughout the imaging area.  Although 
designed to primarily investigate the large-scale structure of the universe, the imaging component 
of the SDSS has obtained high-quality multicolor photometry for about $10^8$ stellar objects in the 
Milky Way, which represents the largest and most homogeneous database on Galactic stellar 
populations ever obtained.  A notable feature of this database is that it was compiled in a new 
photometric system consisting of five unique passbands ($\uprime$, $\gprime$, $\rprime$, $\iprime$, 
and $\zprime$) that were specifically designed for the SDSS to provide continuous coverage over the 
$entire$ optical wavelength range \citep{Fukugita1996}.

While the SDSS was the first to implement the standard $\ugrizprime$ photometric system, analogous 
versions of these same filters are also currently in use with CCD imagers installed on the Gemini 
Telescopes, the Canada-France-Hawaii Telescope, and the Hubble Space Telescope.  In addition, the 
very fact that the SDSS has already provided such a large database of photometry of stars and 
galaxies implies that this photometric system is also becoming widely accepted as the filter set of 
choice for many planned ground-based observational projects and large-scale sky surveys (e.g., 
LSST, OmegaCam, Pan-STARRS, VST).  Despite this, much of our current observational and theoretical 
knowledge of resolved stellar populations is based largely on the conventional Johnson-Kron-Cousins 
$UBV(RI)_C$ photometric system, with a few other studies relying on niche systems like the 
Str\"{o}mgren, DDO, Vilnius, and Geneva systems.  Consequently, the empirical and theoretical tools 
that would tie the standard $\ugrizprime$ system to the fundamental properties of observed stellar 
populations have yet to be defined.  Specifically, neither well-calibrated fiducial stellar 
population sequences nor reliable color--$\Teff$ relations are currently available, and yet, 
without them, it is impossible to fully exploit the capabilities of the SDSS data set as well as 
complementary studies employing these same filters.

In order to remedy these deficiencies, there is good reason to rely on $\ugrizprime$ observations 
of star clusters within our own Galaxy.  Clusters are the ideal stellar population templates 
because, despite a handful of exceptions (e.g., $\omega\,$Cen and M$\,$22), their constituents are 
believed to be effectively coeval, equi-distant, and nearly identical in terms of their heavy 
elemental abundances.  As a result, their color-magnitude diagrams (CMDs) generally exhibit 
extremely tight and well-populated sequences of stars that span several orders of magnitude in 
brightness.  Their wide distribution in metallicity is also suitable for characterizing how the 
photometric properties of stellar populations vary as a function of $\FeH$.  Consequently, cluster 
observations offer the perfect data sets to define fiducial stellar population sequences that cover 
a broad range of stellar parameter space.  These sequences serve as a set of empirical 
``isochrones" that not only facilitate the analysis of other stellar populations data, but also 
provide calibrators for stellar evolutionary models that are transformed to the observed CMDs via 
theoretically-derived color-$\Teff$ relations (e.g., see \citealt{Brown2005}).  Given the fact that 
the standard $\ugrizprime$ system was introduced only a short time ago, however, the photometric 
database for star clusters remains too small to accomplish the tasks mentioned above.

Unfortunately, the SDSS alone cannot provide a sufficient database since (1) the main Sloan survey 
with the 2.5$\,$m telescope is based on instrumental $ugriz$ passbands that are very similar to, 
but not quite identical to, the $\ugrizprime$ passbands with which the $standard$ Sloan photometric 
system was defined on the 1.0$\,$m telescope of the USNO Flagstaff Station 
\citep{S02}.\footnote{\citet{Tucker2006} have published simple linear transformation equations that 
relate the instrumental $ugriz$ magnitudes to standard $\ugrizprime$ with reasonable precision 
(2--3\%).  At some time in the future more sophisticated transformations may be derived that will 
enable users to predict standard-system indices from the main-survey observations with a high level 
of accuracy, at least for normal stellar spectral-energy distributions.  Our own instrumental 
observations from the CFHT have been similarly transformed to the standard $\ugrizprime$ system 
rather than to the $ugriz$ main-survey system; see below, and \citet{Paper1}.}  In addition, (2) 
the brightest stars lying on the red-giant branches (RGBs) of the nearest clusters saturate during 
$\sim60\,$s drift-scan exposure times, (3) the SDSS imaging data do not extend deep enough to 
provide good photometric precision for some of the fainter stars lying on the main sequences in the 
more distant metal-poor globular clusters, (4) the survey footprint does not reach down to the 
Galactic plane where the majority of metal-rich open clusters reside.  These reasons imply that the 
SDSS may not be the most ideal source of star cluster photometry for the derivation of fiducial 
stellar population sequences.

Hence, we recently began an extensive observational project aimed at thoroughly exploring the 
nature of stellar populations in the $\ugrizprime$ photometric system via observations of Galactic 
star clusters.  Our first paper \citep[][hereafter Paper I]{Paper1} presented a network of fainter 
secondary standard stars for the $\ugrizprime$ system in selected star cluster fields to better aid 
observers on large-aperture telescopes in calibrating their photometry.  In fact, the current 
investigation relies on these standards to calibrate a sample of high-quality and homogeneous 
$\ugrizprime$ photometry obtained on the 3.6$\,$m Canada-France-Hawaii Telescope (CFHT).  This 
photometry is subsequently employed here to derive a set of accurate stellar population fiducial 
sequences that span a broad range in both magnitude and metallicity.  In our third and final paper 
(Clem et al. 2008, in preparation) we will utilize these fiducials to test a new grid of 
theoretical color--$\Teff$ relations and bolometric corrections for the $\ugrizprime$ system that 
have been calculated from synthetic spectra.

The following sections present the details related to the observation, reduction, and compilation 
of the photometry collected at the CFHT that will be used to derive a set of fiducial stellar 
population sequences for the $\ugrizprime$ system.  In Section 2 we describe the observational 
setup employed at the CFHT to collect the cluster photometry as well as the data reduction 
procedure, including the important step of calibrating the observed cluster photometry to the 
standard $\ugrizprime$ system.  Section 3 presents the details of the fiducial sequence derivation 
process.  Finally, a short summary of our results, and a discussion of the usefulness of these 
fiducials for stellar populations research is given in Section 4.

\section{The CFHT Star Cluster Survey}

\begin{figure}
\plotone{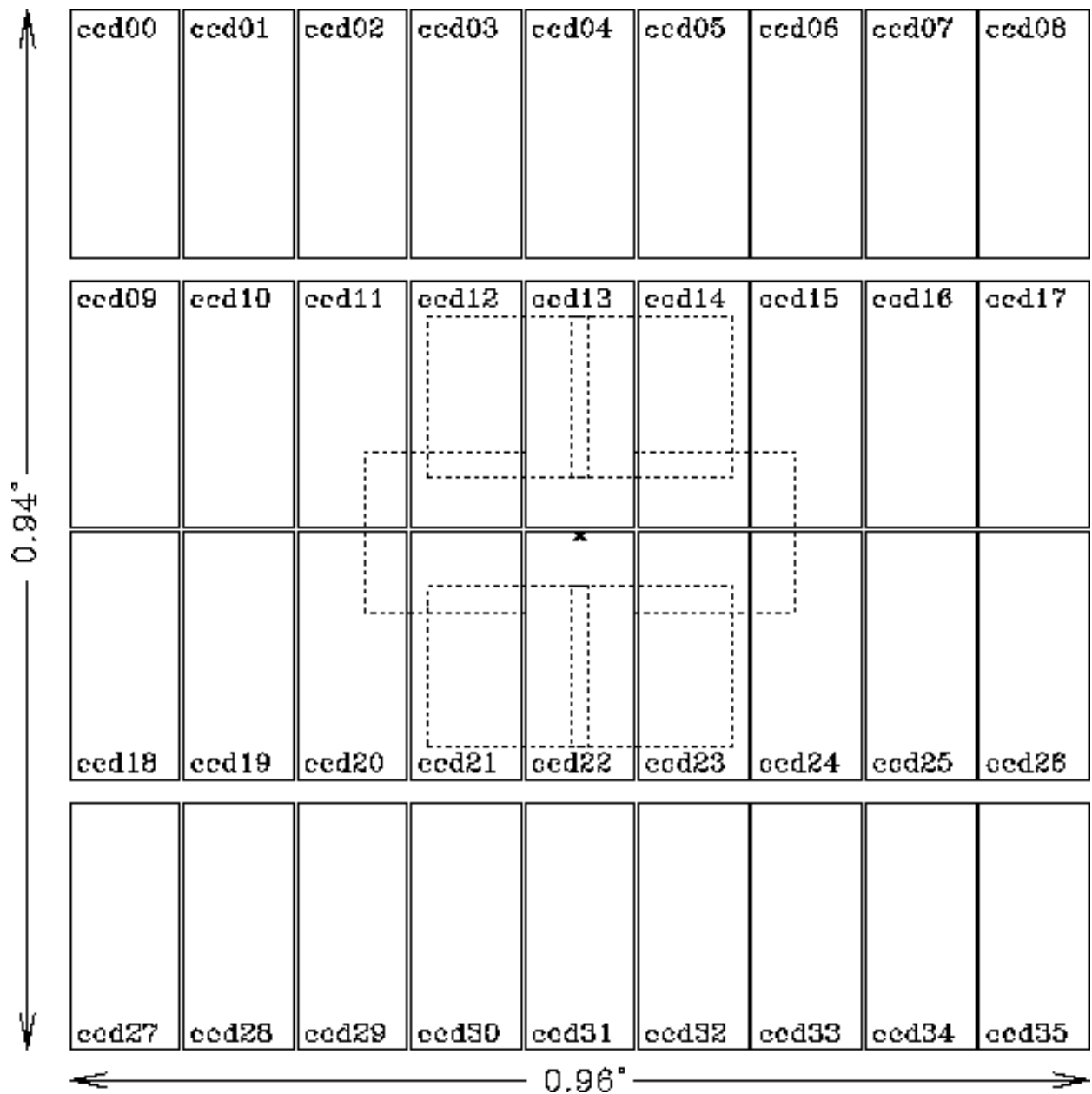}
\caption{Schematic showing the layout of the 36 individual CCD chips in the MegaCam mosaic 
camera.  Each chip measures 2048$\times$4612 pixels and projects to $\sim$6.4$\times$14.4$\arcmin$ 
at the CFHT f/4 prime focus resulting in a full field of $\sim$0.96$\times$0.94 degrees for the 
entire mosaic.  The cross near the center of the mosaic (located $\sim14\arcsec$ below the top of 
ccd22) corresponds to the location of the optical axis of the telescope in the focal plane.  The 
boxes denoted by dotted lines indicate the approximate locations of the 6 fields containing the 
secondary $\ugrizprime$ standards (cf. Figure 2 in Paper I).}
\label{fig:figure01}
\end{figure}

To address the need for fiducial sequences in the $\ugrizprime$ system, an observational program 
aimed at obtaining high-quality photometry for a number of Galactic star clusters was conducted on 
the CFHT in early 2004.  One of the most notable features of these cluster observations is the fact 
that they were obtained using CFHT's wide-field mosaic imager known as ``MegaCam".  As shown in 
Figure \ref{fig:figure01}, MegaCam contains 36 individual CCDs that combine to offer nearly a full 
1x1 degree field of view with high angular resolution ($\sim0.187\arcsec$ pixel$^{-1}$ at the f/4 
prime focus).  Moreover, MegaCam operates with a set of $\gprime\rprime\iprime\zprime$ filters 
whose effective wavelengths and bandwidths are very similar to those of the USNO/SDSS.  For 
observations in the UV, however, a slightly different filter than $\uprime$ is employed; this 
so-called $\ustar$ filter was designed to take advantage of the superb sensitivity of the MegaCam 
CCDs at short wavelengths along with the reduced atmospheric extinction in the UV at high 
altitudes.  

\begin{figure}
\plotone{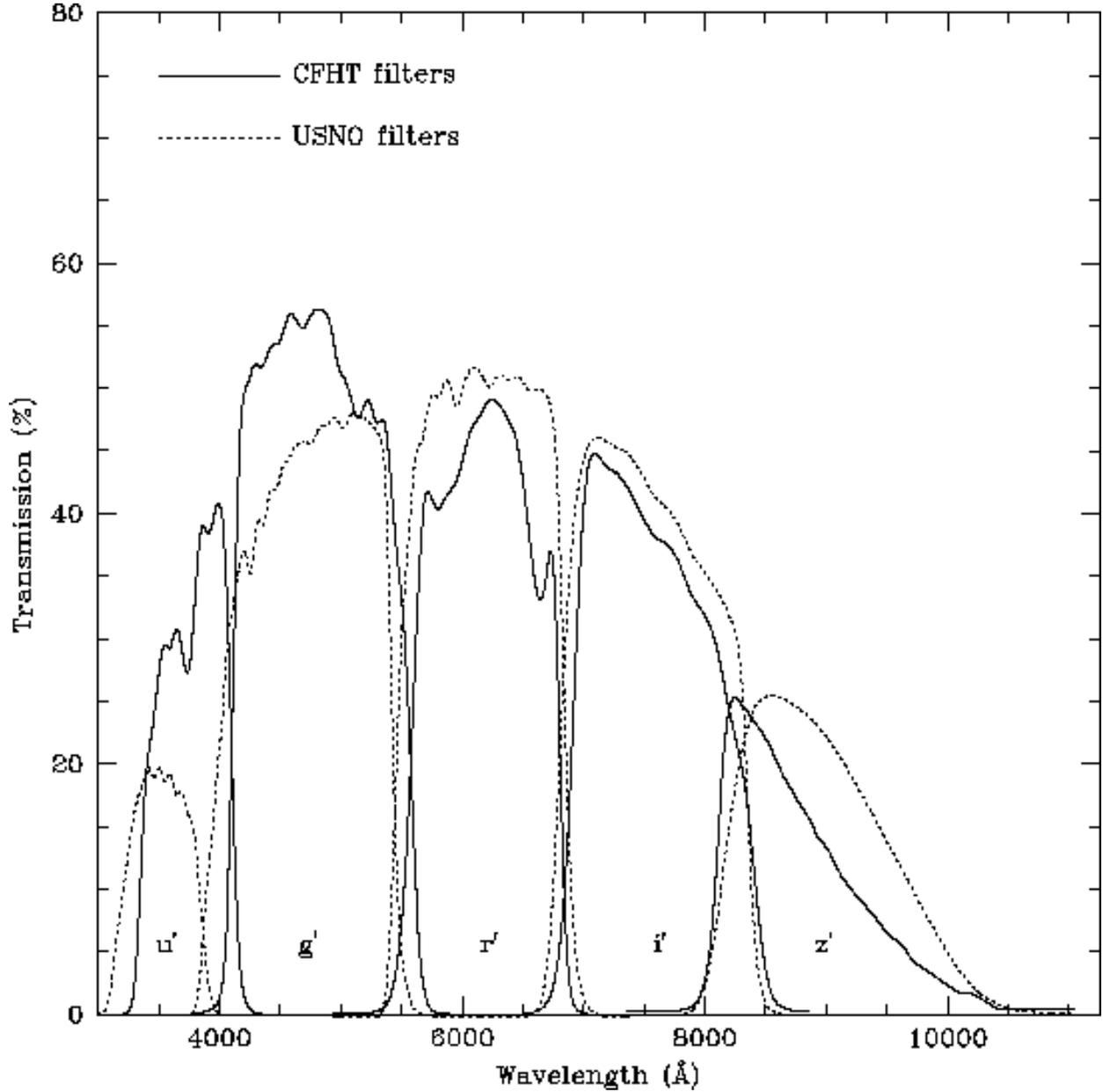}
\caption{Spectral coverages of the CFHT/MegaCam $\cfhtugriz$ and USNO $\ugrizprime$ filter 
sets.  Note that the effective wavelength of MegaCam's $u^{*}$ filter is shifted redward by about 
200\rm{\AA} compared to the $\uprime$ filter.}
\label{fig:figure02}
\end{figure}

In order to graphically compare the USNO and MegaCam photometric systems, we present in Figure 
\ref{fig:figure02} the spectral coverages of both filter sets.  Note that both sets of response 
functions presented in the figure are a result of convolving the raw filter profiles with the 
reflection/transmission characteristics of the telescope optics and the quantum efficiency of the 
detectors employed at both the USNO and CFHT telescopes\footnote{Electronic versions of tables 
containing this information for the USNO are available at 
http://www-star.fnal.gov/ugriz/Filters/response.html while analogous data for CFHT's MegaCam can be 
found at http://www.cfht.hawaii.edu/Instruments/Imaging/Megacam/specsinformation.html}.  The 
differences between $\ustar$ and $\uprime$ mentioned above are clearly evident in the plot with the 
central wavelength of the $\ustar$ filter positioned about $200\rm{\AA}$ $redder$ than that of 
standard $\uprime$ filter employed at the USNO.  As discussed later in this investigation, this 
fact poses a particular problem in transforming photometry observed in MegaCam's $\ustar$ filter to 
$\uprime$ on the standard system.  Apart from the differences between $\ustar$ and $\uprime$, 
however, the agreement between the remaining four filters seems quite good.  The only exceptions 
are the two $\zprime$ filters where the USNO version appears to have more response towards longer 
wavelengths.  This difference can be explained by the fact that the both the MegaCam and USNO 
$\zprime$ filters are manufactured to have no long wavelength cutoff \citep[see][]{Fukugita1996}.  
Instead, their redward edges are defined by the long-wavelength quantum efficiency characteristics 
of the detector employed.

\subsection{Observations and Data Reduction}

Table \ref{tab:table1} presents a list of the dates when data for the program clusters were 
collected on the CFHT during the 2004A observing semester.  The observing run identifications 
provided in the second column denote blocks of several consecutive nights when the same 
instrumental setup was in place on the telescope, and all raw science images collected during these 
blocks were preprocessed using the same run-averaged master bias and flat-field frames.  It is 
important to note that, due to the nature of the ``queue-scheduled" mode of observing operations at 
CFHT, the cluster data were collected on nights when actual sky conditions at the telescope closely 
matched the tolerances specified in the initial project proposal (i.e., near photometric conditions 
during dark or gray time with moderately good seeing).  As a result, the observations were 
generally conducted on non-consecutive nights, and a complete set of cluster observations in all 
five filters may not have been collected on the same night or even during the same observing run 
(for example, the M$\,$3 data were collected over 4 separate nights spanning 3 different observing 
runs).

Our goals for this project were to obtain a series of five short- and five long-exposure images in 
each filter for each cluster in our target list.  This not only ensured a good signal-to-noise 
ratio for stars extending from the tip of the RGB down to a few magnitudes below the turnoff in 
each filter, but also allowed a star to be detected multiple times to help improve the precision of 
its final photometry.  Furthermore, the telescope was dithered by a few tens of arcseconds between 
exposures to allow detection of stars that may have fallen on gaps in the MegaCam mosaic in 
previous frames.  In total, 287 separate MegaCam images in the $\cfhtugriz$ filters was collected 
over the 12 separate nights listed in Table \ref{tab:table1}.  All but one of these nights were 
deemed photometric on the basis of the observing logs, observer's notes, and weather conditions at 
the Mauna Kea site with atmospheric seeing conditions ranging between $0.54\arcsec$ and 
$1.45\arcsec$ (FWHM; median of $\sim0.93\arcsec$) over all nights.  Table \ref{tab:table2} lists 
the number of short and long exposure images obtained per filter on a cluster-by-cluster basis.  It 
is worthwhile to note that that in addition to the clusters listed in Table \ref{tab:table2}, the 
globular cluster M$\,$5 and the open cluster M$\,$67 were included as targets in the observing 
proposal.  However, M$\,$67 was not observed during this period, while only 8 $\gprime$ frames were 
taken for M$\,$5.  As a result, these two clusters are excluded from consideration for the 
remainder of the analysis.  In addition, the 8 $\iprime$ images for M$\,$3 obtained during 
non-photometric conditions were also excluded from the data reductions.  This left 271 MegaCam 
images remaining to be processed in the analysis below.

It is important to note that all of the raw science images acquired for this investigation were 
preprocessed by CFHT's Elixir project \citep{Elixir} prior to their distribution to the principal 
investigators.  This involved the standard steps of overscan correction, bias subtraction, 
flat-fielding (using run-averaged twilight sky flats), masking of bad pixels, and fringe removal 
from the $\iprime$ and $\zprime$ images.  The Elixir project also provided a preliminary 
astrometric calibration and photometric analysis for each MegaCam image.  The latter involved a 
normalization of the background level in each CCD to enforce a nearly identical instrumental 
zero-point for all chips and ensure that the final processed data show only small variations from a 
constant background over the entire mosaic.

As a result of the processing done by Elixir, the data analysis could proceed directly to the 
extraction of the instrumental PSF photometry from each MegaCam image.  In this respect, no attempt 
was made prior to these reductions either to combine the 36 different CCD images into a single 
master exposure or to co-add the dithered exposures, but rather the digital images from the 36 
individual CCDs were processed separately as if they came from distinct cameras.  This was 
advantageous since the act of assembling multiple CCDs into a large mosaiced image often requires 
the ``resampling" of the individual images to account for subtle chip-to-chip differences in scale 
and rotation.  This type of processing generally requires pixel interpolations and extrapolations 
that can distort the PSF in certain regions of the mosaic and may lead to spurious photometry for 
some objects during the PSF fitting.

Since the reduction of the CFHT images was conducted in a manner similar to that described in Paper 
I, only a brief review is provided here.  First, a UNIX shell script was employed to 
non-interactively run the PSF-building and star subtraction DAOPHOT/ALLSTAR routines 
\citep{Stetson1987, StetsonHarris1988} on each individual CCD frame.  After star lists and 
associated instrumental PSF photometry for each frame had been derived, the correction of the 
relative profile-fitted photometry to a more absolute, aperture-based system was accomplished with 
concentric aperture growth curves derived using the DAOGROW package \citep{Stetson1990}.  Finally, 
the geometric transformations of the natural ($x$, $y$) coordinate system of the individual CCD 
images to an astrometrically meaningful system ($\xi$, $\eta$; see, e.g., \citealt{Smart1965}) 
based on the USNOB-1.0 catalog \citep{Monet2003} was accomplished via a set of third-order 
polynomials.  

\subsection{Photometric Calibrations}

As a result of the efforts described in Paper I, most star clusters in the CFHT survey contain a 
sizable number of local standard stars whose magnitudes have been referred to the $\ugrizprime$ 
system with an accuracy of 1\% or better in each filter.  Therefore, calibrating the cluster 
photometry relies solely on comparing the observed instrumental magnitudes for these stars to their 
counterparts on the standard system in order to solve for the transformation coefficients using 
robust least-squares analysis.  In the beginning stages of the calibration process, each CCD 
exposure from the mosaic was treated separately.  That is, the transformation constants were 
allowed to be determined freely and independently on the basis of the local standard stars 
contained in the image.  Unfortunately, since the range in air mass spanned by the cluster 
observations on any given night was typically quite small ($\Delta(sec\,z)\lesssim0.2$), it was 
soon discovered that the derived extinction coefficients often took on negative values and/or were 
wildly inconsistent between the different CCD images.  Due to these findings, the canonical 
atmospheric extinction coefficients for Mauna Kea (as determined by the Elixir project) were 
employed for the calibrations rather than having them computed from the data.  That is, all chips 
are assigned $K_{u^{*}}=0.35$, $K_{\gprime}=0.15$, $K_{\rprime}=0.10$, $K_{\iprime}=0.04$, and 
$K_{\zprime}=0.03\,$mag per air mass for all photometric nights.

With the extinction coefficients set to constant values for all chips, secondary runs through the 
calibrations were performed with one less unknown in the transformation equations.  Upon 
completion, a weighted average of the linear color terms was then calculated and imposed as an 
additional known constant common to all chips for a final calibration run that left only the 
photometric zero points to be recomputed on a chip-by-chip basis.  Given the fact that the spectral 
response of a telescope/detector system is largely determined by the combined effects of the 
transmission of the atmosphere, the reflectivity and transmissivity of the telescope components, 
and throughput of the filters, it can be expected that chip-to-chip differences in the computed 
color terms vary by only a few percent (provided that the CCDs are similar in design and the filter 
is spatially uniform).  Indeed, from our own derivation of separate color terms for different 
chips, we found that the largest variations in the color coefficients between the different chips 
from any single photometric night amounted to $\sim4\%$ in the $\uprime$ filter.

\begin{figure}
\plotone{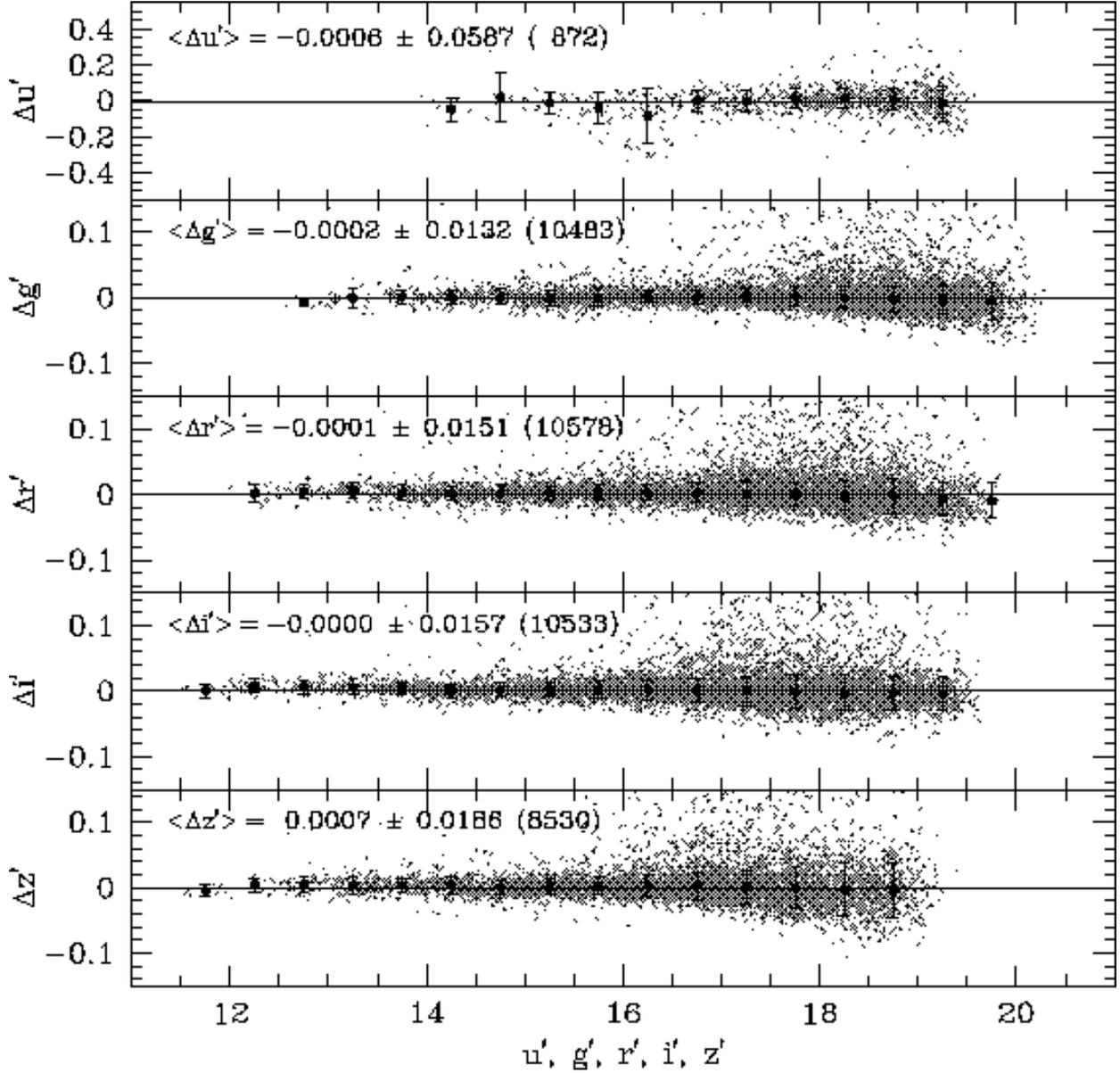}
\caption{Plot of the photometric differences for stars observed over the 11 photometric nights on 
CFHT that also appear in the database of secondary $\ugrizprime$ standards derived in Paper I.  
Each $\Delta$mag is plotted against its respective magnitude on the standard system.  The gray 
dots represent individual secondary standard stars in common between the two data sets while the 
large black circles designate the median difference in bins of 0.5 mag.  The error bars associated 
with the median values correspond to a robust measure of the dispersion in the differences within 
each bin.  The mean $\Delta$mag differences, associated standard deviations, and number of stars 
used to define the mean are denoted in each panel.}
\label{fig:figure03}
\end{figure}

\begin{figure}
\plotone{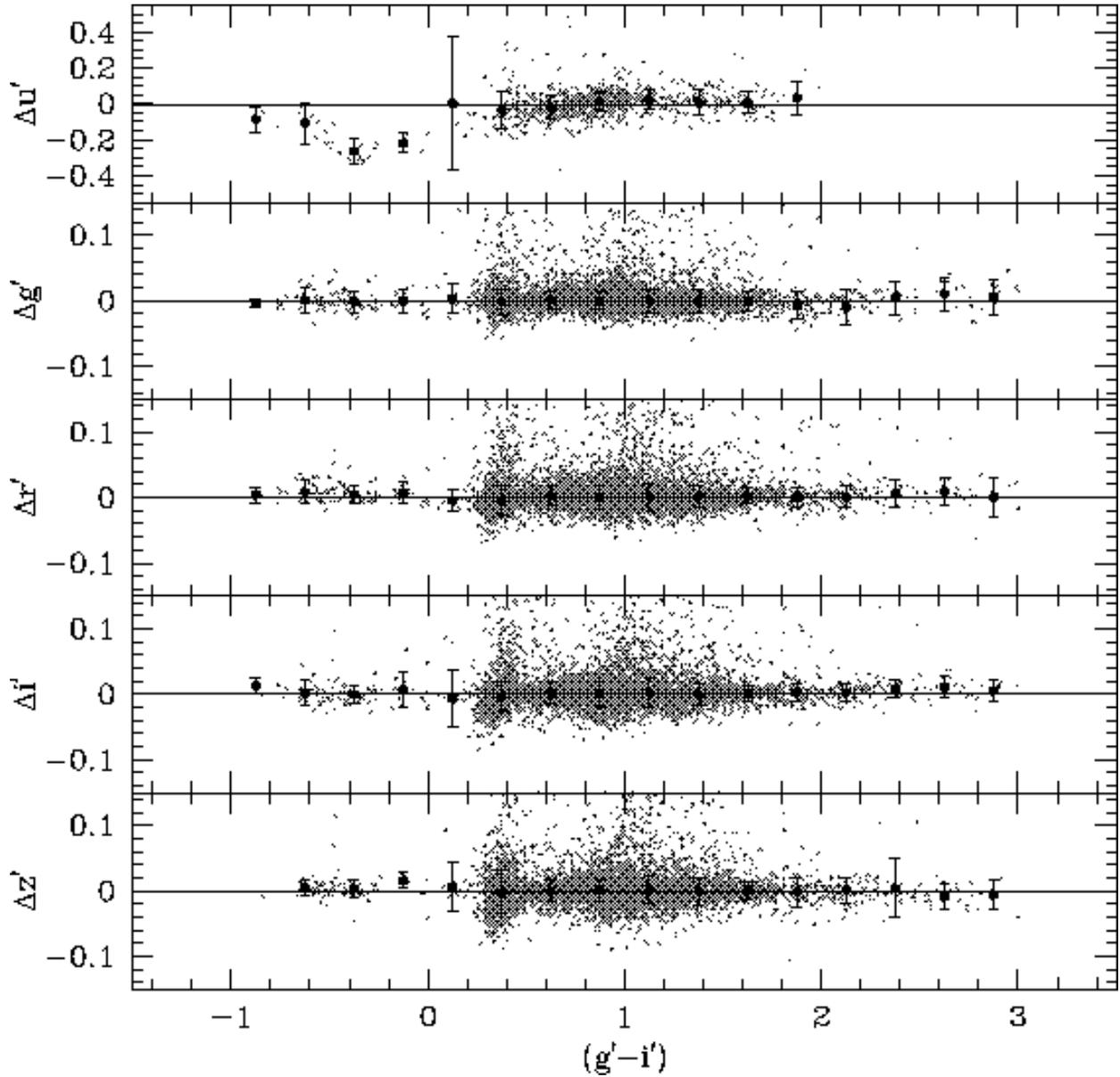}
\caption{Same as Figure \ref{fig:figure03} except the magnitude residuals are plotted against the 
standard ($\gprime-\iprime$) color.  Each large black circle represents the median difference in 
bins of 0.25 mag in color.  Note the strong deviations in $\Delta \uprime$ as a function of color 
in the top panel.}
\label{fig:figure04}
\end{figure}

Once the transformation constants from all nights were derived, the best estimates of the 
calibrated magnitudes for each secondary standard star observed over all 11 photometric nights 
were then determined.  Figures \ref{fig:figure03} and \ref{fig:figure04} illustrate the extent to 
which the cluster data have been placed on the standard system by comparing the $\ugrizprime$ 
photometry to the mean calibrated photometry for the secondary standards stars.  The small, gray 
data points in each panel represent the difference between the magnitudes on the standard 
$\ugrizprime$ system and the $mean$ magnitude as a function of their respective magnitudes in 
Figure \ref{fig:figure03} and $(\gprime-\iprime)$ color in Figure \ref{fig:figure04}.  To better 
aid in the detection of any systematic trends not immediately evident to the eye, each large dot 
represents the unweighted median difference for stars in intervals of 0.5 in magnitude or 
0.25$\,$mag in $(\gprime-\iprime)$ color with their corresponding error bars providing a robust 
measure of the spread in the residuals within each bin (i.e., ($\pi/2)^{1/2}$ times the mean 
absolute deviation).

A few features in both figures warrant detailed explanations.  First, there is a noticeable excess 
of scatter in the data points towards positive values in each panel of Figure \ref{fig:figure03}; 
this seems to indicate that a sizable fraction of the secondary standards with $\rprime\gtrsim15$ 
have had their magnitudes on the standard system measured too bright compared to the values derived 
here.  The most likely explanation for this effect is the fact that these standards are located 
closer to cluster cores and hence their photometry in the DAO images has been contaminated by light 
from nearby stars.  This would serve to make the secondary standards appear brighter than they 
actually are.  As mentioned in Paper I, since atmospheric seeing conditions averaged around 
$3-4\arcsec$ at the DAO for the establishment of the secondary standards, it is reasonable to 
expect a higher probability of spurious photometry resulting from the effects of crowding in the 
DAO images compared to the CFHT observations (with average seeing values of $\sim1.0\arcsec$).  To 
better test this statement, we have produced plots similar to those in Figures \ref{fig:figure03} 
and \ref{fig:figure04} where stars are culled based on how crowded they are in the DAO images as 
derived by $sep$ index (a detailed definition of this parameter is presented below).  While we do 
not present these plots here, we did find that the more crowded stars tended to exhibit more 
scatter towards positive residual values as well as larger means and standard deviations when 
compared to plots that considered only less crowded stars.  Hence, we strongly feel that our 
derived photometric transformations are sufficient to transform the observed MegaCam photometry to 
the standard system based on the fact that the median magnitude differences tend to lie within 
0.01$\,$mag of zero difference over the entire ranges in $\gprime$, $\rprime$, $\iprime$, and 
$\zprime$, and the lines corresponding to zero difference tend to pass through the densest parts of 
the point distributions.

Secondly, the situation with the $\uprime$ magnitude differences plotted as a function of color in 
the top panel of Figure \ref{fig:figure04} appears quite troublesome.  In particular, there is 
considerable disagreement in the residuals towards bluer and redder colors which would seem to 
suggest that the use of a single linear color term in the transformation equations is inadequate to 
account for the bandpass mismatch between the $u^{*}$ and $\uprime$ filters.  As mentioned above, 
CFHT's $u^{*}$ filter was constructed to be substantially different than the $\uprime$ filter in 
order to take advantage of the good UV sensitivity of MegaCam and reduced atmospheric extinction at 
short wavelengths atop Mauna Kea. The effective wavelength of the $u^{*}$ filter is about 
200\rm{\AA} redder than that of $\uprime$, and, as shown in Figure \ref{fig:figure02}, this places 
most of the $u^{*}$ response $redward$ of the Balmer discontinuity at 3700\rm{\AA}; this has 
profound implications for transforming the observed $u^{*}$ magnitudes for B- and A-type stars 
(i.e., those with the largest Balmer jumps) to the standard system.

\subsection{Transforming $\ustar$ to $\uprime$}

\begin{figure}
\plotone{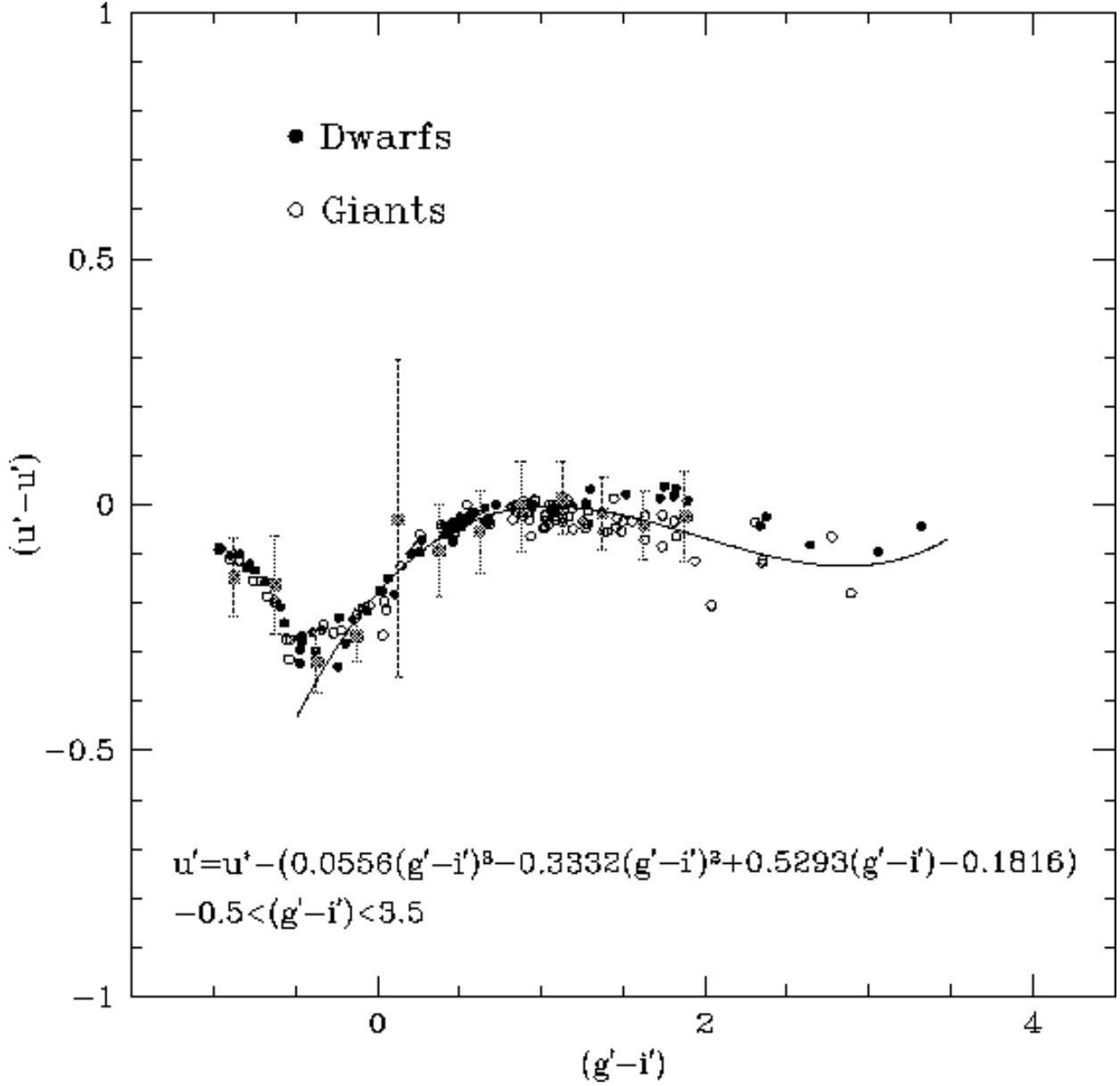}
\caption{Differences between MegaCam's $\ustar$ and standard $\uprime$ magnitudes as computed by 
convolving the respective filter transmission functions shown in Figure \ref{fig:figure02} with 
the stellar spectral energy distributions as presented by \citet{GunnStryker1983}.  Giant and 
dwarf stars in the Gunn \& Stryker sample are denoted by open and solid circles, respectively.  
The median differences in the observed residuals from the top panel of Figure \ref{fig:figure04} 
are also plotted as solid gray squares to illustrate the agreement between our own data and the 
computed magnitudes.  The solid line provides the third-order polynomial that was fit to the data 
to help transform $\ustar$ to $\uprime$.}
\label{fig:figure05}
\end{figure}

The deviations shown in the top panel of Figure \ref{fig:figure04} for $\Delta \uprime$ appear to 
indicate that the transformation of MegaCam's $\ustar$ to $\uprime$ is more complex than can be 
accounted for using a simple linear color term.  As a result, we have endeavored to derive a more 
realistic higher-order polynomial transformation that would better convert MegaCam's $\ustar$ 
photometry to $\uprime$ on the standard system.  To keep this derivation as empirically-based as 
possible, we employed our own photometric data as well as the colors and magnitudes computed by 
convolving the filter transmission functions for both the standard $\ugrizprime$ and MegaCam's 
$\cfhtugriz$ systems (c.f. Figure \ref{fig:figure02}) with the spectral energy distributions for 
real stars.  For the latter, we opted to use the SEDs presented in the \citet{GunnStryker1983} 
spectrophotometric catalog that cover a wide range in luminosity and spectral class in order to 
better investigate the difference between $\ustar$ and $\uprime$ photometry for different stellar 
types.  When the differences between the ``computed" $\ustar$ and $\uprime$ magnitudes are compared 
against $(\gprime-\iprime)$, as shown in Figure \ref{fig:figure05}, we see that indeed there is a 
rather complex behavior in the residuals that appears to coincide quite well with the residuals 
shown in the top panel of Figure \ref{fig:figure04} (the gray squares with error bars denote the 
same median points plotted in Figure \ref{fig:figure04}).  While it would appear that the dwarfs 
and giants follow a slightly different trend towards redder colors (i.e., 
$(\gprime-\iprime)\gtrsim1.0$), we have decided to find a single, multi-order function that would 
best correct the $\ustar$ photometry for most stellar types.  Consequently, the solid line plotted 
in the figure indicates the third-order polynomial we have fitted to both sets of data to better 
define the transformation between $\ustar$ and $\uprime$.  Note that this fit is only valid over 
the range $-0.5<(\gprime-\iprime)<3.5$ and is not intended to correct the $\ustar$ photometry for 
extremely blue or red stars.

\begin{figure}
\plotone{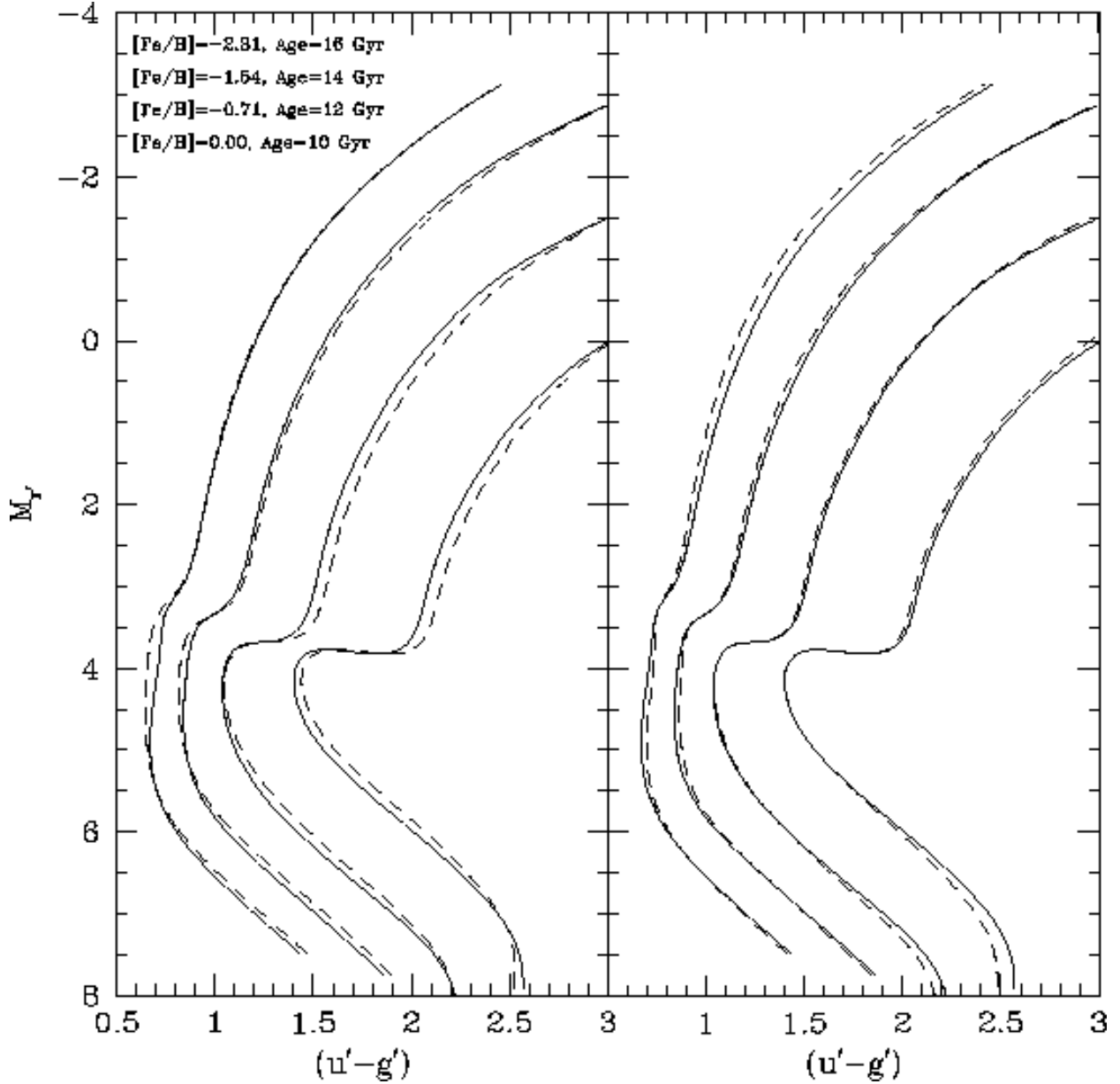}
\caption{Comparisons between theoretical isochrones, with denoted ages and metallicities plotted 
left-to-right, as transformed to the observed [$(\uprime-\gprime)$, $M_{\rprime}$] plane using 
color-temperature relations derived from ATLAS9 synthetic spectra for both MegaCam's $\cfhtugriz$ 
(dashed lines) and the standard $\ugrizprime$ (solid lines) filter transmission functions.  The 
left-hand panel compares the two sets of isochrones using the simple linear color term initially 
derived to transform MegaCam's $\ustar$ photometry to the standard system, while the right-hand 
panel employs the polynomial transformation denoted in Figure \ref{fig:figure05}.  Note the 
better overall agreement between the two sets when the third-order transformation is used.}
\label{fig:figure06}
\end{figure}

To test the quality of this new non-linear transformation between $\ustar$ and $\uprime$, Figure 
\ref{fig:figure06} presents a comparison between various \citet{VandenBerg2006} isochrones with the 
denoted ages and metallicities (as plotted left to right in the figure) that have been transformed 
to the [$(\uprime-\gprime)$, $M_{\rprime}$] plane using color-temperature relations computed from 
the ATLAS9 synthetic spectra presented by \citet{Castelli1997}.  The left-hand panel employs the 
linear color terms that were initially derived to transform the MegaCam photometry to the standard 
system, and it shows that the two sets of isochrones are in considerable disagreement in a number 
of different locations due to the mismatch between the $\ustar$ and $\uprime$ filters, most notably 
in the turnoff region for metal-poor stars, and the RGB and main sequence regions for more 
metal-rich stars.  The right-hand panel, on the other hand, plots isochrones where $\ustar$ has 
been corrected using the third-order function shown in Figure \ref{fig:figure05}.  Reassuringly, 
the new nonlinear transformation seems to provide theoretical loci in the region 
$0.0\gtrsim(\uprime-\gprime)\gtrsim3.0$ that are in quite good agreement with those on the standard 
system , and, with a few exceptions discussed below, both sets of isochrones overlap each other to 
within 0.01 mag over the entire range in magnitude.  This fact bodes well for our derivation of the 
fiducial cluster sequences in $\uprime$ since the stars that define RGB, SGB, turnoff, and upper-MS 
loci for all clusters in the survey fall within this color range.

In regards to the few remaining obvious differences between the two theoretical loci in the 
right-hand panel for the upper-RGB for the most metal-poor isochrone and the lower-MS for the most 
metal-rich, they would appear to indicate that our polynomial relation is unable to correct the 
$\ustar$ magnitudes for $all$ types of stars (in particular, those stars that lie at the extremes 
of luminosity, color, and/or metallicity).  Unfortunately, as far as we are aware, there does not 
exist a sufficient database of stars with varying metallicities, temperatures, and/or luminosities 
that have been observed in both the $\ustar$ and $\uprime$ filters to conduct a more rigorous 
investigation into the differences between these two filters for a wide-variety of stellar types.  
Also, it is important to stress, that the transformations between $u^{*}$ and $\uprime$ that we 
have derived here are invalid for blue horizontal branch stars or other hot stars, such as blue 
post-AGB stars, sdB stars, white dwarfs, or some blue stragglers.  Although we are quite confident 
that the fiducials we derive below are accurately calibrated to the standard system for the 
$\gprime$, $\rprime$, $\iprime$, and $\zprime$ filters, we strongly advise the reader to use 
caution when employing the sequences that include $\uprime$ filter for the interpretation of 
stellar photometry, especially for metal-poor giants or metal rich dwarfs, due to the simple fact 
that the fiducials may not be adequately transformed to $\uprime$ on the standard system for these 
types of stars.

\section{The Cluster Photometry}

The transformation terms computed during the calibration to the standard $\ugrizprime$ system 
described in the previous section were subsequently applied to the instrumental magnitudes that 
were derived for all detected objects in every CCD image for each cluster field.  Simultaneously, 
the zero points between the relative profile-fitting magnitude system and the standard one were 
redetermined by direct comparison to the local secondary standards within each cluster field on a 
frame-by-frame basis.  This final step in the reduction process compensates for uncertainties 
caused by short term fluctuations in the extinction or errors in the aperture corrections.  While 
this does nothing to improve the absolute calibration of the photometry to the standard system in 
the mean, it does improve the frame-to-frame repeatability of the measurements by ensuring that the 
photometry from each image is now referred to a common magnitude zero point defined by the local 
secondary standards.  In addition, the transformation of the natural ($x$, $y$) coordinates of the 
stars in each image to an astrometrically meaningful system based on the USNOB-1.0 catalog 
facilitates the matching of stars from different chips and different exposures and results in a 
single master star list for the entire field surrounding each cluster.  

\subsection{Refining the Sample}

Given that we have on hand approximately 16.9 million individual magnitude and position 
measurements derived for some 650,000 distinct objects in 5 different clusters, it is inevitable 
that a sizable number of these detections will be non-stellar objects (e.g., background galaxies, 
cosmic rays, satellite or meteor trails, etc.) or image blemishes (e.g., defective pixels, 
diffraction spikes, etc.).  Moreover, when dealing with crowded cluster fields such as these, the 
photometry for a significant fraction of the legitimate stars will undoubtedly be contaminated by 
light from neighboring objects even under the most ideal seeing conditions.  As a result, when the 
cluster photometry is plotted on color-magnitude or color-color diagrams for the purpose of 
analysis, these spurious objects and crowded stars may contribute increased scatter or broadening 
of the primary cluster sequences, and it is better to exclude them from consideration when deriving 
the fiducials.  While it is obviously not feasible to censor problematic measurements by hand, the 
various programs that have been used to extract the PSF instrumental photometry from the CCD images 
output certain image-quality and data-reliability indices that can be used to reject spurious 
detections or non-stellar objects from consideration.  In addition to these indices, the discussion 
below describes the mechanics of the so-called ``separation index" \citep[$sep$; 
see][]{Stetson2003} that is quite effective in culling severely crowded stars from the cluster data 
sets.

In brief, the definition of the separation index is based on the fact that the typical seeing 
profile for each star in a particular image is well approximated by the Moffat function 
\citep{Moffat1969}:

\begin{equation}
S(r)\propto\frac{F}{[1+(r/r_o)^2]^{\beta}},
\end{equation}
\vspace{0.05in}

\noindent{where $r$ is the distance from the star's centroid, $r_o$ is some characteristic radius 
that can be related to FWHM of the stellar brightness profile, $F$ is just the stellar flux 
determined from $F\propto10^{-0.4m}$; where $m$ is the apparent magnitude, $S(r)$ is the surface 
brightness of the stellar profile at radius $r$, and $\beta$ is a parameter that governs the shape 
of the stellar profile.  Based on this definition, if one assumes a reasonable value for $\beta$ 
(typically 1.5--2.5 for stellar profiles in digital images) and FWHM for the seeing value, it is a 
simple matter to compute the surface brightness produced by a particular star with both an apparent 
magnitude $m$ and a centroid position at any point in the field.  Based on this definition, the 
$sep$ index for any given star can be mathematically expressed as:}

\begin{equation}
sep_i=\frac{S_i(0)}{\Sigma_{j \ne i} S_{j}(r_j)}.
\end{equation}
\vspace{0.05in}

\noindent{Here $S_{i}(0)$ is the surface brightness at the centroid of the star in question and 
$S_{j}(r_{j})$ is the surface brightness contribution from the $j^{th}$ neighboring star situated 
at a distance $r_j$ away.}

\begin{figure}
\plotone{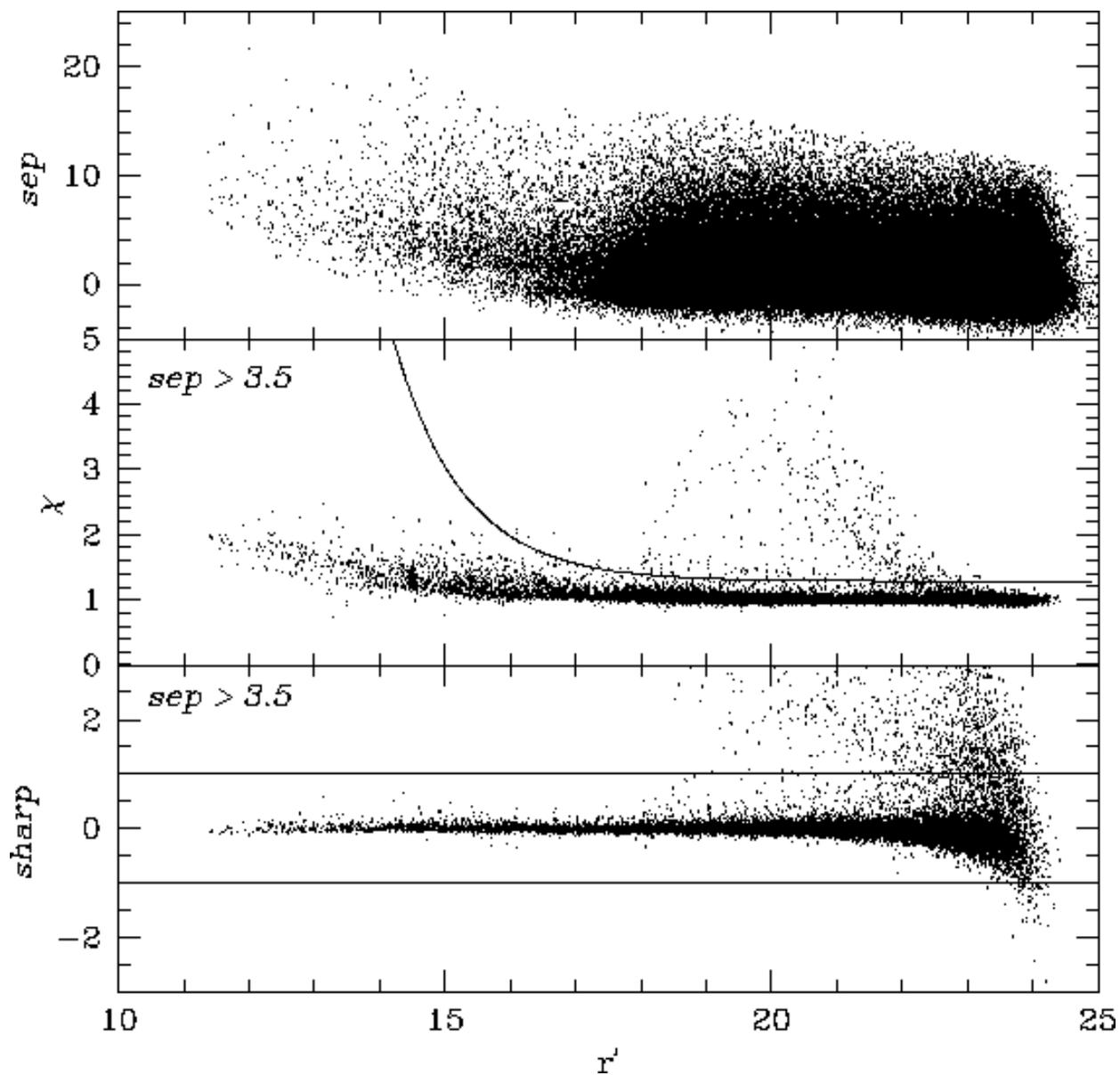}
\caption{Plots of the image-isolation and image-quality indices $sep$, $\chi$, and $sharp$ versus 
apparent $\rprime$ magnitude for stars in the globular cluster M$\,$13.  Only those stars with 
$sep>3.5$ are plotted in the bottom two panels.  Stars lying below the solid curve in the middle 
panel together with those having $-1<sharp<1$ in the bottom panel are retained in the sample for 
the derivation of the cluster fiducial sequences.}
\label{fig:figure07}
\end{figure}

The computation of the $sep$ index for the 5 different cluster data sets assumes the typical values 
of FWHM$\,=\,1.0\arcsec$ and $\beta\,=\,2$ and uses the apparent $\rprime$ magnitude to define the 
fluxes for the individual stars.  In order to save computational time, the determination of $sep$ 
for any particular star in the field considers contributions only from those stars lying within 10 
times the assumed FWHM.  The top panel of Figure \ref{fig:figure07} shows the plot of the derived 
$sep$ index versus apparent $\rprime$ magnitude for stars in M$\,$13.  As evidenced by the higher 
concentration of points at increasing magnitudes, fainter stars are more susceptible to 
contamination by light from neighboring stars in the field than bright ones.  Since the M$\,$13 
turnoff corresponds to $\rprime \sim18.6$, the scattering of points to brighter magnitudes and 
higher $sep$ values primarily correspond to stars lying on the RGB and HB of this cluster.  Based 
on examinations of the cluster CMDs using different $sep$ cuts, it was determined that stars with 
$sep>3.5$ (i.e., stars where the summed wings of all other stars in the field amount to no more 
than $\sim4\%$ of the central surface brightness of the star itself) produced the most well-defined 
cluster sequences.  Therefore, the remainder of this discussion only considers those stars with 
$sep$ values above 3.5.

During the process of deriving PSF magnitudes the DAOPHOT/ALLSTAR software computes two 
image-quality indices known as $\chi$ and $sharp$ for every detected object in a CCD image.  In the 
final reduction of all the data for a particular cluster, the individual $\chi$ and $sharp$ 
measurements for each star are then averaged and reported in the data tables.  Briefly, $\chi$ is 
simply a measure of the agreement between the object's observed brightness profile and the derived 
PSF model (i.e., the quality of the fit between the model PSF and the object).  As shown in the 
middle panel of Figure \ref{fig:figure07}, the $\chi$ values for the vast majority of objects with 
$sep>3.5$ in M$\,$13 tend to cluster around $\chi\approx1$ over the entire magnitude range which 
would indicate that they are legitimate stars.  Those at larger $\chi$ values, on the other hand, 
are most likely either non-stellar objects or stars whose brightness profiles are corrupted by 
image defects or diffraction spikes.  Stars lying above the solid curve shown in the same panel are 
excluded on the basis of the $\chi$ values.  Finally, a plot of the $sharp$ index versus apparent 
$\rprime$ magnitude in the bottom panel of Figure \ref{fig:figure07} shows that real stars have a 
propensity to hover in a narrow range centered on zero.  This is due to the fact that the $sharp$ 
index measures the degree to which an object's intrinsic angular radius differs from that of the 
model PSF.  Therefore, detections with large positive $sharp$ values have larger characteristic 
radii compared to the PSF model and are most likely resolved galaxies, while those with 
significantly negative $sharp$ values have apparent radii smaller than the seeing profile, and thus 
are unlikely to be astronomical objects viewed through the atmosphere and telescope optics; they 
probably correspond to an image blemishes or cosmic rays.  
As a result, one can safely assume that 
objects with $|sharp|<1$ have a high degree of probability of being real stars.

\begin{figure}
\plotone{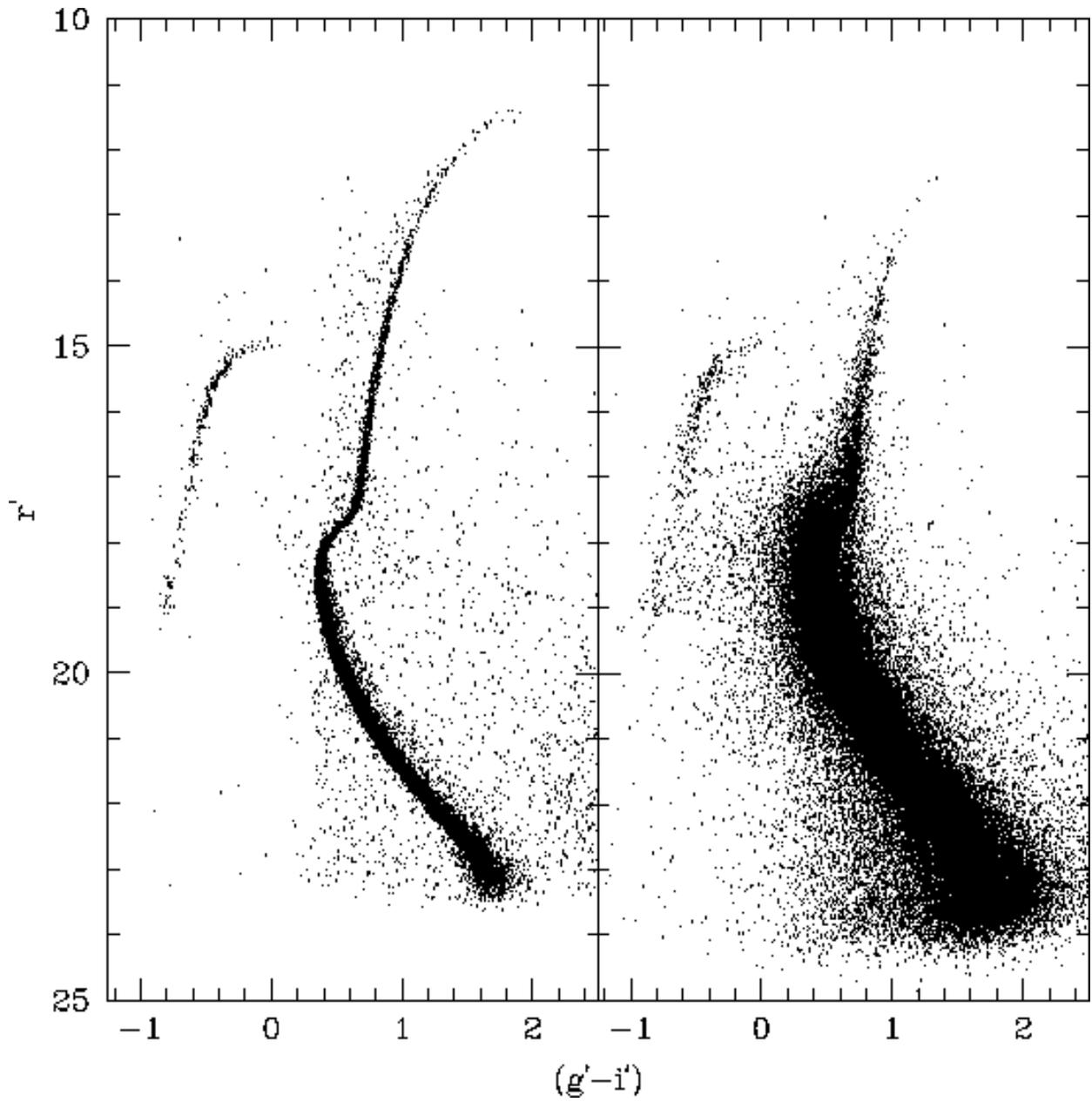}
\caption{Two $(\gprime-\iprime, \rprime)$ CMDs for stars in the field surrounding the globular 
cluster M$\,$13.  The left-hand panel plots those stars judged to have the highest quality 
photometry on the basis of their $sep$, $\chi$, and $sharp$ values as described in the text.  The 
right-hand panel presents stars that are excluded from the deviation of the fiducial sequences due 
to their poorer photometry.  Note the more diffuse nature of the primary cluster sequences in 
the right-hand panel.}
\label{fig:figure08}
\end{figure}

To demonstrate the effectiveness of the $\chi$, $sharp$, and $sep$ indices in culling crowded stars 
and spurious objects from the photometry lists and producing extremely well-defined cluster 
sequences, Figure \ref{fig:figure08} shows two CMDs for M$\,$13 with those stars that survived the 
cuts plotted in the left-hand panel and those that did not in the right.  Note the well-defined and 
very tight cluster sequences extending from the RGB to the lower main sequence in the left-hand 
panel.  In contrast, stars that were excluded in the right-hand panel result in a quite diffuse and 
noisy main sequence, turnoff, and lower-RGB regions due largely to the effects of crowding.  It is
important to mention that while the stars plotted in the left panel do not represent a $complete$ 
sample of all the cluster members, they do provide a suitable $representative$ sample for the 
derivation of the fiducial sequences.

\subsection{Defining the Fiducials}

With objects from each of the cluster data files rejected or accepted according to the cuts in 
$\chi$, $sharp$, and $sep$ mentioned above, the definition of the fiducial sequences from the 
cluster photometry proceeds by defining the ridge lines of the stellar locus in color-magnitude 
space.  Due to the various possible combinations of different colors and magnitudes that are 
available to plot a cluster's CMD, the $\rprime$ magnitude was adopted as the primary ordinate 
against which the median colors were defined since the cluster loci are rarely double valued in 
$\rprime$, and the level of completeness at faint magnitudes is the best for $\rprime$.  
Therefore, each ridge line is created by determining the median color of stars that lie within 
different $\rprime$ magnitude bins.  The size of these bins is arbitrarily adjusted along the 
cluster locus to include a sufficient number of stars to define a median color.  For example, 
larger magnitude bins are defined in parts of the CMD where the photometric scatter is larger at 
the faint end and where the number of stars is scarce at the bright end.  Smaller bins are 
employed for areas of the cluster loci with large curvature and numerous stars (e.g., between the 
turnoff and base of the RGB).  Outlying stars are iteratively clipped during the determination of 
the median color to ensure that the ridge line is not significantly skewed.  While this technique 
seemed to work quite well, there were some regions of the CMD where the number of stars is just 
too small, the scatter in the sequences is too large, or the cluster locus is double valued (i.e., 
the subgiant branch of NGC$\,$6791) for an accurate median color to be defined; in these cases the 
location of the points defining the ridge lines are determined by eye estimation.

\begin{figure}
\plotone{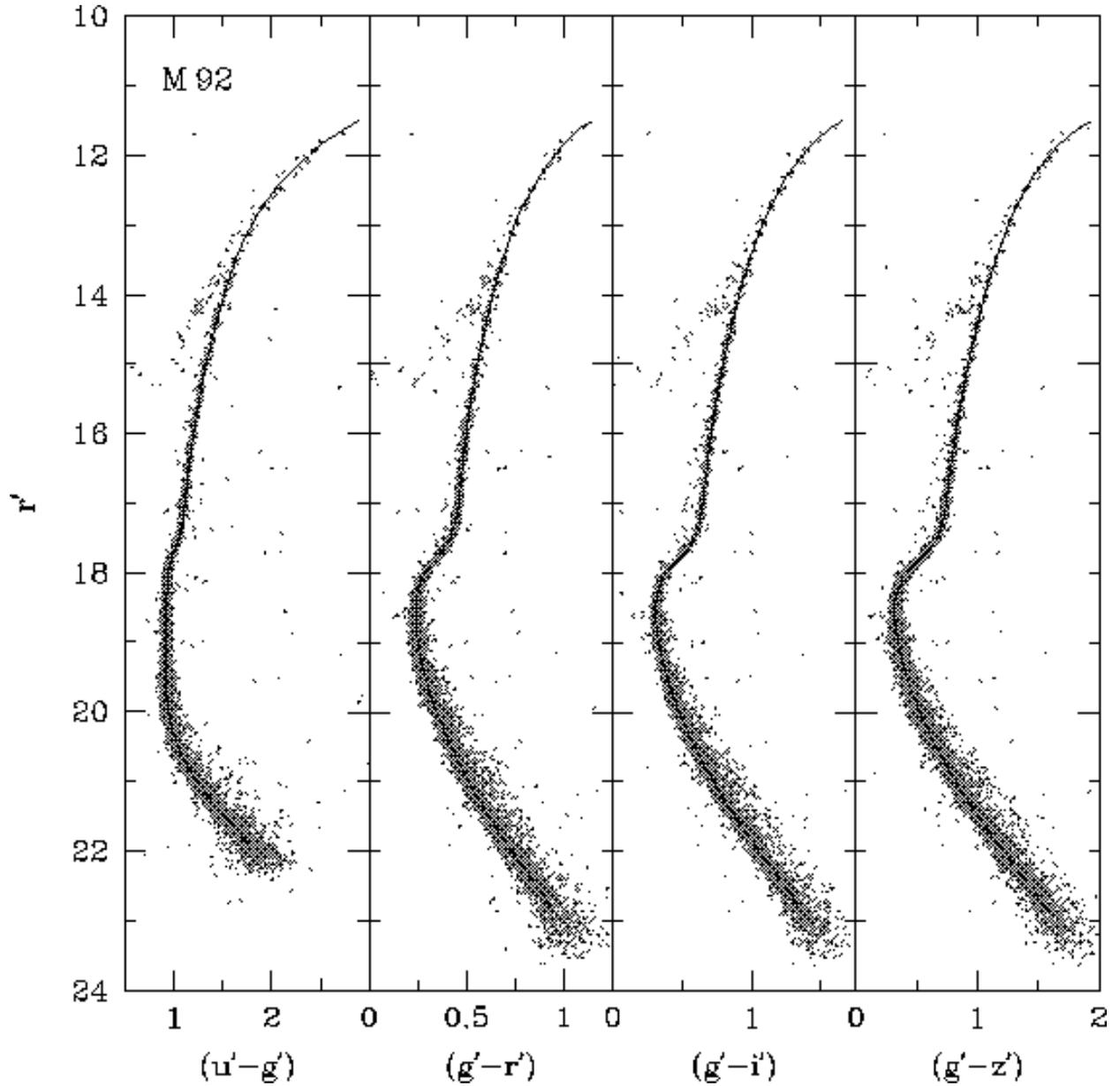}
\caption{Various $\ugrizprime$ CMDs and associated derived fiducial sequences for the globular cluster 
M$\,$92.  Each panel includes only those stars judged to have the highest quality photometry 
based on their values of $\chi$, $sharp$, and $sep$.}
\label{fig:figure09}
\end{figure}

\begin{figure}
\plotone{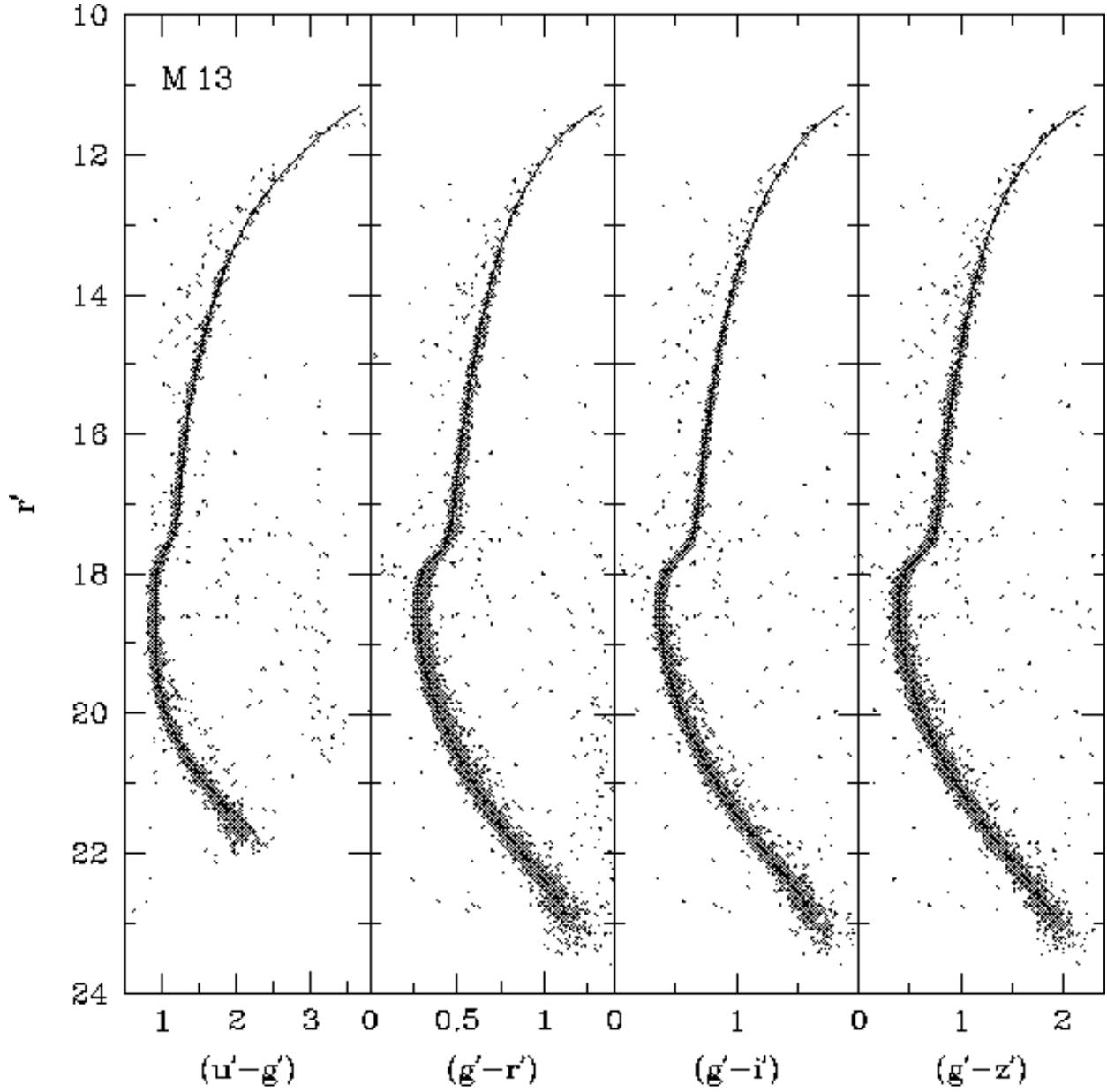}
\caption{Same as Figure \ref{fig:figure09}, but for the globular cluster M$\,$13.}
\label{fig:figure10}
\end{figure}

\begin{figure}
\plotone{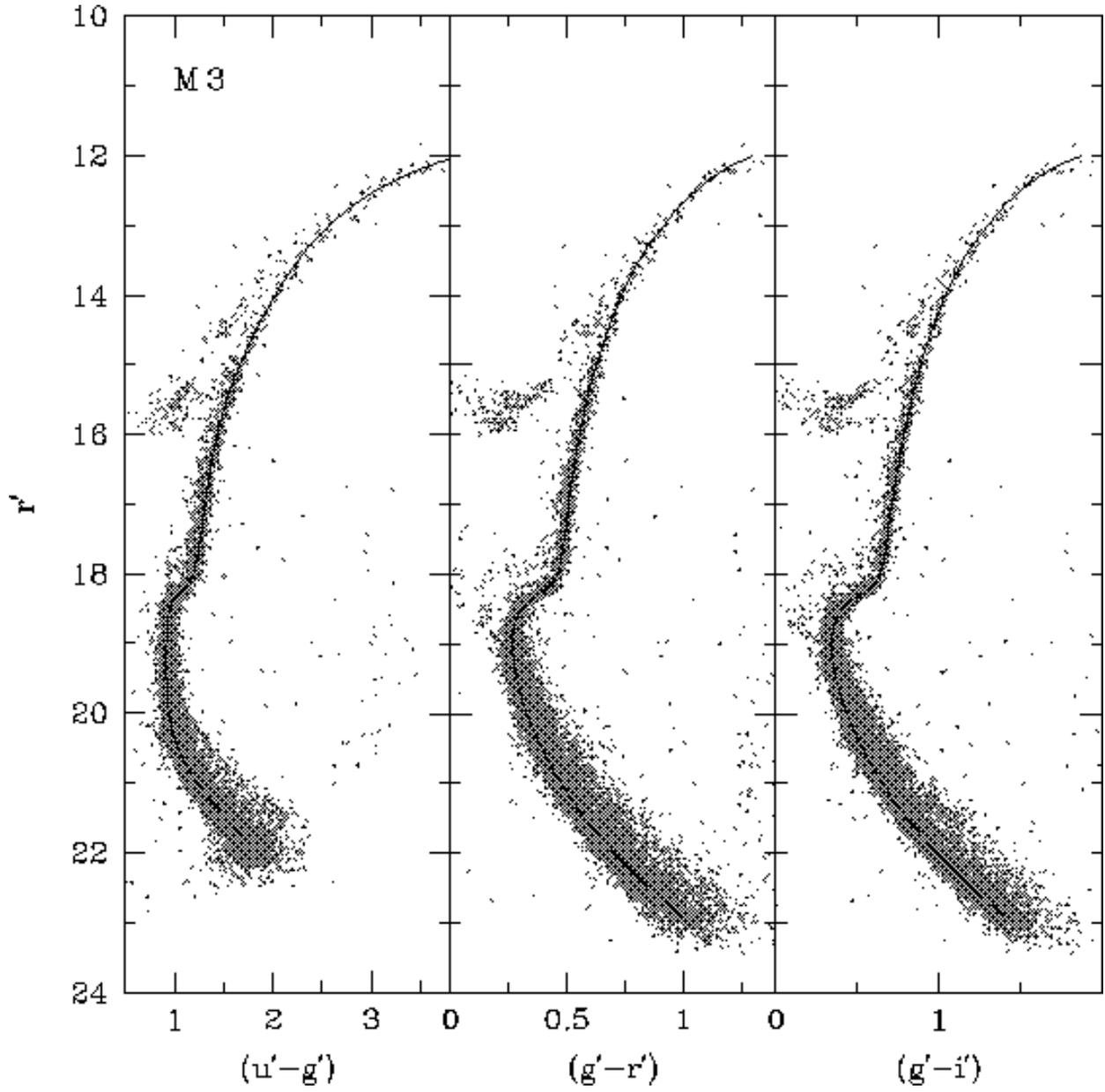}
\caption{Same as Figure \ref{fig:figure09}, but for the globular cluster M$\,$3.  Note that 
$\zprime$ photometry is not available for this cluster.}
\label{fig:figure11}
\end{figure}

\begin{figure}
\plotone{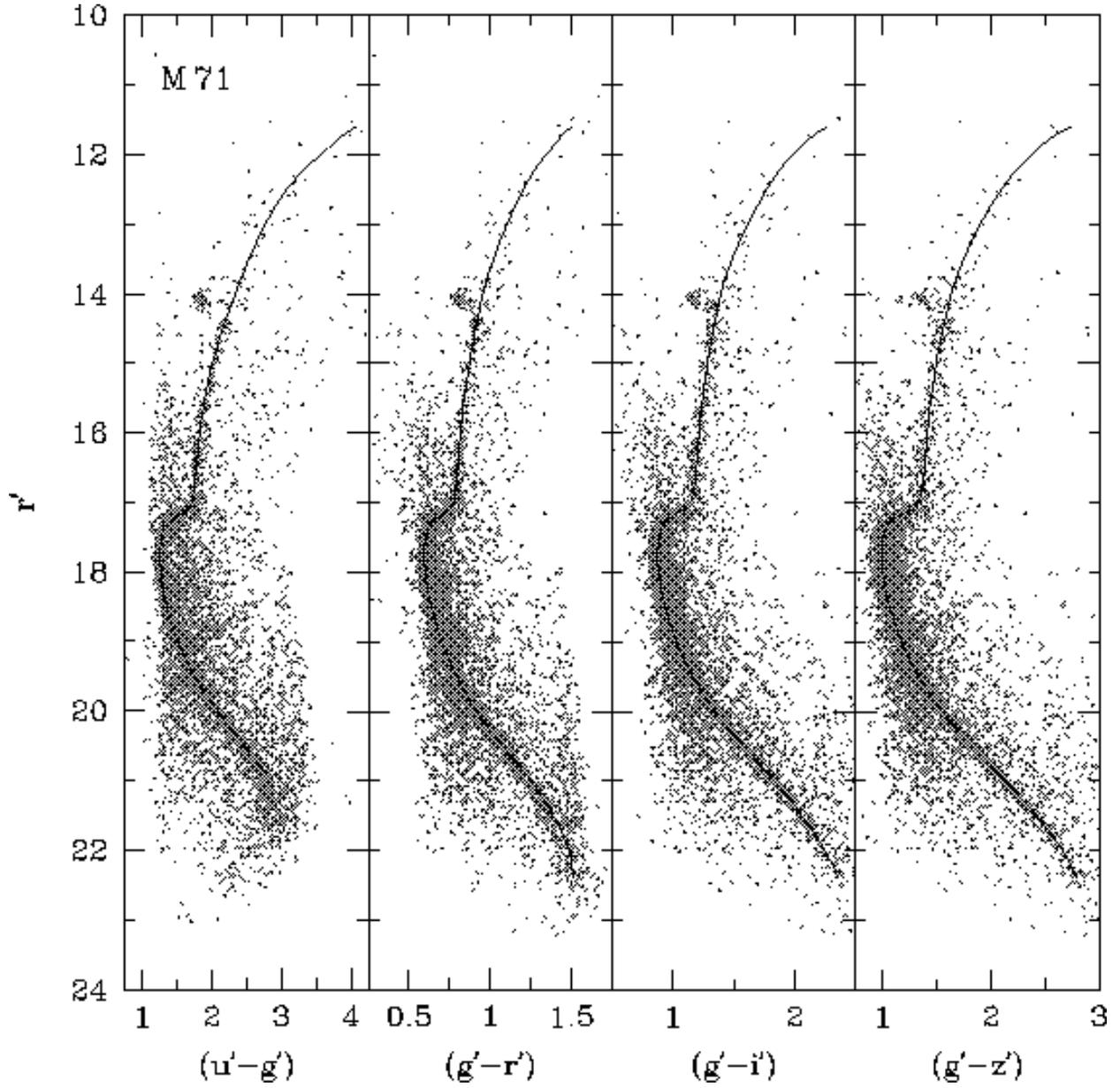}
\caption{Same as Figure \ref{fig:figure09}, but for the globular cluster M$\,$71.  Each 
panel plots only those stars that lie within a radius of $2.5\arcmin$ from the cluster center in 
order to reduce field star contamination in the CMDs.}
\label{fig:figure12}
\end{figure}

\begin{figure}
\plotone{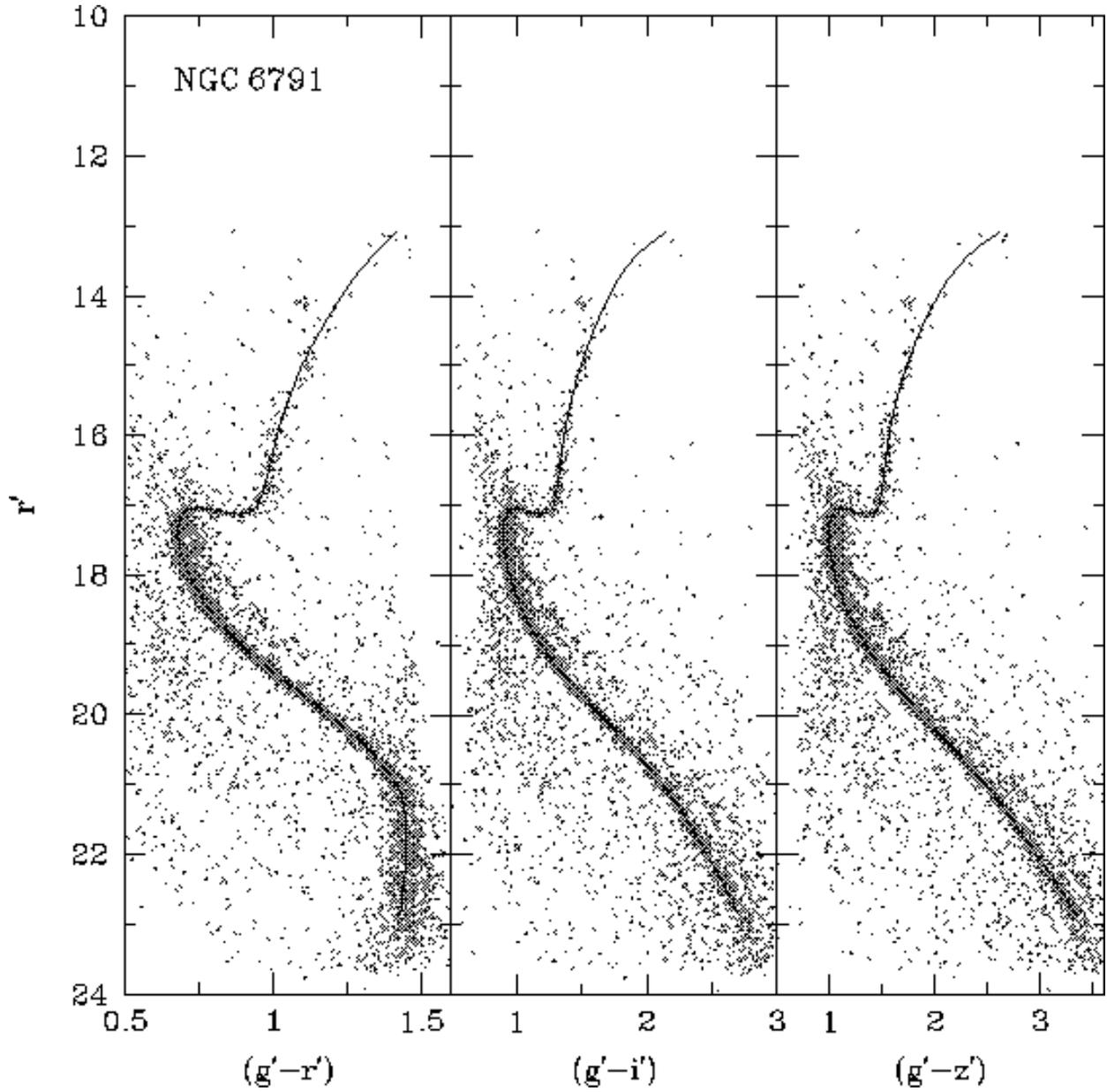}
\caption{Same as Figure \ref{fig:figure09}, but for the open cluster NGC$\,$6791.  $\uprime$ 
photometry is not available for this cluster.  Each panel plots only those stars that lie 
within a radius of $5\arcmin$ from the cluster center in order to reduce field star contamination 
in the CMDs.}
\label{fig:figure13}
\end{figure}

Figures \ref{fig:figure09} through \ref{fig:figure13} present the various CMDs of each cluster in 
the sample along with their associated ridge lines spanning the MS, SGB, and RGB (tabulated in 
Tables \ref{tab:table3}-\ref{tab:table7}).  It is important to note that the photometry for each 
cluster has been censored according to the same $\chi$, $sharp$, and $sep$ cuts mentioned above 
before plotting.  In addition, only those stars that lie within a radius of $2.5\arcmin$ and 
$5.0\arcmin$ of the centers of M$\,$71 and NGC$\,$6791, respectively, have been plotted to help 
reduce field star contamination in their CMDs.  These imposed cuts appear to have been quite 
successful in yielding extremely well-defined and tight loci of stars extending from the upper-RGBs 
down to approximately 4 magnitudes below the turnoff points.

\section{Summary}

Using high-quality, homogeneous observations obtained with the wide-field MegaCam imager on the 
3.6m Canada-France-Hawaii Telescope, we have derived fiducial stellar population sequences for the 
Galactic star clusters M$\,$92, M$\,$13, M$\,$3, M$\,$71, and NGC$\,$6791 in the $\ugrizprime$ 
photometric system.  These sequences, which span a wide range in both metallicity and magnitude, 
have been accurately calibrated to within 1\% of the standard $\ugrizprime$ system using a set of 
secondary standard stars derived in Paper I.  As a result of our efforts, we anticipate that these 
fiducial sequences will serve as valuable tools for the interpretation of other stellar population 
investigations involving the $\ugrizprime$ bandpasses by virtue of the fact that they represent a 
set of empirical isochrones for both metal-poor and metal-rich stars having wide-ranging physical 
parameters.  Indeed, a preliminary set of the fiducials presented in this work has already been 
employed in the interpretation of CMDs for a number of newly discovered Milky Way satellites (see, 
for example, \citealt{Belokurov2006, Belokurov2007a, Belokurov2007b}).

In addition to the usefulness of these fiducial sequences for the interpretation of observed data, 
they also provide an excellent test of the accuracy of color-temperature relations and bolometric 
corrections that have been derived from model atmospheres and synthetic spectra.  In a future paper 
(J.~L. Clem et al. 2008, in preparation) we intend to perform tests of such synthetic color and 
magnitude transformations by comparing isochrones models to the fiducials derived here.  Our aim is 
to assess the quality of the color-temperature relations and bolometric corrections in providing 
isochrones that can reproduce the observed cluster photometry when reasonable estimates of the 
cluster reddening, metallicity, and distance are assumed, as well as test their consistency when 
isochrone fits to the $\ugrizprime$ photometry are compared to those in other photometric systems 
(e.g., BV(RI)$_C$ and $uvby$).

It is important for the reader to note, however, that the fiducials derived in this investigation 
are presented on the $\ugrizprime$ system and $not$ on the natural photometric system of the 
2.5$\,$m SDSS survey telescope (i.e., the $ugriz$ system).  Subtle differences exist between the 
two systems such that photometry reported on both systems for an identical stellar sample can 
differ systematically by as much as a few hundredths of a magnitude in some filters.  Therefore, we 
caution against using these fiducials to interpret and/or analyze $ugriz$ photometry from the SDSS 
without first applying appropriate transformation relations (see \citealt{Tucker2006}).  Although 
these transformations may not be appropriate for stars with strong emission features in their 
spectra or stars with extreme colors (i.e., later than M0 spectral class), we expect they are good 
enough to transform the fiducials presented here on the $\ugrizprime$ system to the SDSS 2.5$\,$m 
$ugriz$ system, while keeping the uncertainties in the photometric zero points on the AB system 
\citep[see][]{OkeGunn1983} to within a few percent.

\acknowledgements

This work has been supported by an Operating Grant to D. A. V. from the Natural Sciences and 
Engineering Research Council of Canada.  This paper was prepared with support from NSF grant AST 
05-03871 to A.~U.~Landolt.

\clearpage
\newpage
\begin{deluxetable}{lcccccccc}
\tablecaption{Observing Log for the CFHT Star Cluster Survey}
\tablewidth{0pt}
\tablehead{\colhead{UT Dates}           &
           \colhead{Run ID}             &
           \colhead{$u^{*}$}            &
           \colhead{$\gprime$}          &
           \colhead{$\rprime$}          &
           \colhead{$\iprime$}          &
           \colhead{$\zprime$}          &
           \colhead{Clusters Observed}  &
           \colhead{Photometric?}       }
\startdata
2004-05-13  & 04AM04 &      7  &      0  &      0  &      0  &      0  & M$\,$92              & Y   \\
2004-05-23  & 04AM05 &     20  &      0  &      0  &      0  &     10  & M$\,$3, M$\,$92      & Y   \\
2004-06-10  & 04AM06 &      0  &     10  &     10  &     10  &      0  & M$\,$92              & Y   \\
2004-06-11  &        &      5  &     10  &     10  &     10  &      0  & M$\,$3               & Y   \\
2004-06-14  &        &      0  &      0  &      0  &      8  &      0  & M$\,$3               & N   \\
2004-06-19  &        &      0  &     10  &     10  &     10  &      0  & NGC$\,$6791          & Y   \\
2004-07-07  & 04AM07 &      8  &      6  &      0  &      0  &      0  & M$\,$13, NGC$\,$6791 & Y   \\
2004-07-08  &        &      0  &     10  &     10  &     10  &      0  & M$\,$71              & Y   \\
2004-07-10  &        &     10  &      0  &      0  &      0  &     12  & NGC$\,$6791          & Y   \\
2004-07-13  &        &     10  &      0  &      0  &      0  &     10  & M$\,$71              & Y   \\
2004-07-16  &        &     10  &     10  &     10  &     10  &     10  & M$\,$13              & Y   \\
2004-07-17  &        &      3  &      8  &      0  &      0  &      0  & M$\,$3, M$\,$5       & Y   \\ \tableline
Totals      &        &     73  &     64  &     50  &     58  &     42  &                      &     \\
\enddata
\label{tab:table1}
\end{deluxetable}

\clearpage

\clearpage
\newpage
\begin{deluxetable}{lccccc}
\tablecaption{Number of $u^{*}\gprime\rprime\iprime\zprime$ observations per cluster}
\tablewidth{0pt}
\tablehead{\colhead{Cluster}   &
           \colhead{$u^{*}$}   &
           \colhead{$\gprime$} &
           \colhead{$\rprime$} &
           \colhead{$\iprime$} &
           \colhead{$\zprime$} }
\startdata
M$\,$92     &     17  &     10  &     10  &     10  &     10  \\
M$\,$13     &     10  &     16  &     10  &     10  &     10  \\
M$\,$3      &     18  &     10  &     10  &     18  &      0  \\
M$\,$71     &     10  &     10  &     10  &     10  &     10  \\
NGC$\,$6791 &     18  &     10  &     10  &     10  &     12  \\ \tableline
Totals      &     73  &     64  &     50  &     58  &     42  \\
\enddata
\label{tab:table2}
\end{deluxetable}

\clearpage

\clearpage
\newpage
\begin{deluxetable}{ccccc}
\tablecaption{Ridge lines for the globular cluster M$\,$92.}
\tabletypesize{\scriptsize}
\tablewidth{0pt}
\tablehead{\colhead{$\uprime$}  &  
           \colhead{$\gprime$}  &  
           \colhead{$\rprime$}  &  
           \colhead{$\iprime$}  &  
           \colhead{$\zprime$}  }  
\startdata
 15.550 & 12.650 & 11.500 & 10.992 & 10.710 \\ 
 15.336 & 12.952 & 12.000 & 11.589 & 11.363 \\ 
 15.362 & 13.335 & 12.500 & 12.139 & 11.946 \\ 
 15.541 & 13.746 & 13.000 & 12.670 & 12.495 \\ 
 15.809 & 14.184 & 13.500 & 13.200 & 13.040 \\ 
 16.143 & 14.634 & 14.000 & 13.724 & 13.575 \\ 
 16.502 & 15.093 & 14.500 & 14.245 & 14.107 \\ 
 16.876 & 15.554 & 15.000 & 14.762 & 14.633 \\ 
 17.275 & 16.023 & 15.500 & 15.278 & 15.159 \\ 
 17.690 & 16.497 & 16.000 & 15.790 & 15.681 \\ 
 18.120 & 16.976 & 16.500 & 16.300 & 16.200 \\ 
 18.560 & 17.458 & 17.000 & 16.811 & 16.720 \\ 
 18.735 & 17.648 & 17.200 & 17.016 & 16.928 \\ 
 18.896 & 17.828 & 17.400 & 17.221 & 17.140 \\ 
 19.014 & 17.995 & 17.600 & 17.438 & 17.368 \\ 
 19.108 & 18.145 & 17.800 & 17.668 & 17.614 \\ 
 19.220 & 18.287 & 18.000 & 17.898 & 17.867 \\ 
 19.374 & 18.453 & 18.200 & 18.119 & 18.101 \\ 
 19.549 & 18.638 & 18.400 & 18.329 & 18.320 \\ 
 19.740 & 18.835 & 18.600 & 18.529 & 18.522 \\ 
 19.939 & 19.038 & 18.800 & 18.725 & 18.716 \\ 
 20.147 & 19.248 & 19.000 & 18.919 & 18.907 \\ 
 20.359 & 19.460 & 19.200 & 19.112 & 19.095 \\ 
 20.576 & 19.675 & 19.400 & 19.302 & 19.280 \\ 
 20.799 & 19.894 & 19.600 & 19.493 & 19.464 \\ 
 21.031 & 20.117 & 19.800 & 19.682 & 19.646 \\ 
 21.274 & 20.343 & 20.000 & 19.869 & 19.824 \\ 
 21.947 & 20.916 & 20.500 & 20.331 & 20.264 \\ 
 22.732 & 21.509 & 21.000 & 20.787 & 20.693 \\ 
 23.618 & 22.110 & 21.500 & 21.237 & 21.113 \\ 
 24.516 & 22.729 & 22.000 & 21.683 & 21.527 \\ 
 \nodata & 23.356 & 22.500 & 22.126 & 21.933 \\ 
 \nodata & 23.973 & 23.000 & 22.572 & 22.345 \\ 
\enddata
\label{tab:table3}
\end{deluxetable}

\clearpage

\clearpage
\newpage
\begin{deluxetable}{ccccc}
\tablecaption{Ridge lines for the globular cluster M$\,$13.}
\tabletypesize{\scriptsize}
\tablewidth{0pt}
\tablehead{\colhead{$\uprime$}  &  
           \colhead{$\gprime$}  &  
           \colhead{$\rprime$}  &  
           \colhead{$\iprime$}  &  
           \colhead{$\zprime$}  }  
\startdata
 16.300 & 12.630 & 11.300 & 10.760 & 10.411 \\ 
 15.925 & 12.871 & 11.800 & 11.340 & 11.077 \\ 
 15.830 & 13.233 & 12.300 & 11.902 & 11.690 \\ 
 15.863 & 13.624 & 12.800 & 12.447 & 12.266 \\ 
 16.025 & 14.050 & 13.300 & 12.980 & 12.815 \\ 
 16.279 & 14.494 & 13.800 & 13.508 & 13.355 \\ 
 16.584 & 14.947 & 14.300 & 14.032 & 13.889 \\ 
 16.921 & 15.407 & 14.800 & 14.550 & 14.416 \\ 
 17.282 & 15.870 & 15.300 & 15.065 & 14.942 \\ 
 17.672 & 16.338 & 15.800 & 15.576 & 15.460 \\ 
 18.089 & 16.809 & 16.300 & 16.086 & 15.977 \\ 
 18.520 & 17.286 & 16.800 & 16.594 & 16.491 \\ 
 18.777 & 17.574 & 17.100 & 16.900 & 16.800 \\ 
 18.937 & 17.761 & 17.300 & 17.105 & 17.007 \\ 
 19.063 & 17.939 & 17.500 & 17.312 & 17.223 \\ 
 19.117 & 18.090 & 17.700 & 17.542 & 17.479 \\ 
 19.160 & 18.222 & 17.900 & 17.778 & 17.736 \\ 
 19.292 & 18.385 & 18.100 & 17.994 & 17.965 \\ 
 19.465 & 18.569 & 18.300 & 18.203 & 18.181 \\ 
 19.655 & 18.764 & 18.500 & 18.406 & 18.385 \\ 
 19.861 & 18.970 & 18.700 & 18.605 & 18.583 \\ 
 20.075 & 19.179 & 18.900 & 18.801 & 18.774 \\ 
 20.293 & 19.390 & 19.100 & 18.995 & 18.965 \\ 
 20.523 & 19.607 & 19.300 & 19.187 & 19.150 \\ 
 20.764 & 19.829 & 19.500 & 19.378 & 19.336 \\ 
 21.017 & 20.052 & 19.700 & 19.567 & 19.519 \\ 
 21.290 & 20.278 & 19.900 & 19.755 & 19.699 \\ 
 21.902 & 20.745 & 20.300 & 20.124 & 20.050 \\ 
 22.777 & 21.343 & 20.800 & 20.579 & 20.477 \\ 
 23.764 & 21.963 & 21.300 & 21.027 & 20.886 \\ 
 \nodata & 22.604 & 21.800 & 21.462 & 21.279 \\ 
 \nodata & 23.265 & 22.300 & 21.891 & 21.667 \\ 
 \nodata & 23.906 & 22.800 & 22.324 & 22.059 \\ 
 \nodata & 24.449 & 23.300 & 22.784 & 22.473 \\ 
\enddata
\label{tab:table4}
\end{deluxetable}

\clearpage

\clearpage
\newpage
\begin{deluxetable}{ccccc}
\tablecaption{Ridge lines for the globular cluster M$\,$3.}
\tabletypesize{\scriptsize}
\tablewidth{0pt}
\tablehead{\colhead{$\uprime$}  &  
           \colhead{$\gprime$}  &  
           \colhead{$\rprime$}  &  
           \colhead{$\iprime$}  &  
           \colhead{$\zprime$}  }  
\startdata
 17.220 & 13.300 & 12.000 & 11.420 & \nodata \\ 
 16.588 & 13.558 & 12.500 & 12.054 & \nodata \\ 
 16.490 & 13.918 & 13.000 & 12.617 & \nodata \\ 
 16.566 & 14.310 & 13.500 & 13.162 & \nodata \\ 
 16.751 & 14.733 & 14.000 & 13.695 & \nodata \\ 
 17.003 & 15.175 & 14.500 & 14.219 & \nodata \\ 
 17.290 & 15.626 & 15.000 & 14.739 & \nodata \\ 
 17.611 & 16.084 & 15.500 & 15.255 & \nodata \\ 
 17.985 & 16.555 & 16.000 & 15.770 & \nodata \\ 
 18.386 & 17.028 & 16.500 & 16.285 & \nodata \\ 
 18.810 & 17.507 & 17.000 & 16.797 & \nodata \\ 
 19.244 & 17.987 & 17.500 & 17.308 & \nodata \\ 
 19.418 & 18.179 & 17.700 & 17.511 & \nodata \\ 
 19.573 & 18.367 & 17.900 & 17.716 & \nodata \\ 
 19.695 & 18.546 & 18.100 & 17.929 & \nodata \\ 
 19.714 & 18.679 & 18.300 & 18.159 & \nodata \\ 
 19.752 & 18.806 & 18.500 & 18.403 & \nodata \\ 
 19.898 & 18.975 & 18.700 & 18.618 & \nodata \\ 
 20.077 & 19.164 & 18.900 & 18.824 & \nodata \\ 
 20.274 & 19.365 & 19.100 & 19.025 & \nodata \\ 
 20.481 & 19.571 & 19.300 & 19.224 & \nodata \\ 
 20.693 & 19.782 & 19.500 & 19.417 & \nodata \\ 
 20.916 & 19.997 & 19.700 & 19.609 & \nodata \\ 
 21.144 & 20.214 & 19.900 & 19.799 & \nodata \\ 
 21.387 & 20.436 & 20.100 & 19.988 & \nodata \\ 
 21.641 & 20.660 & 20.300 & 20.176 & \nodata \\ 
 21.920 & 20.889 & 20.500 & 20.364 & \nodata \\ 
 22.696 & 21.473 & 21.000 & 20.824 & \nodata \\ 
 23.587 & 22.072 & 21.500 & 21.277 & \nodata \\ 
 24.491 & 22.698 & 22.000 & 21.720 & \nodata \\ 
 \nodata & 23.350 & 22.500 & 22.154 & \nodata \\ 
 \nodata & 24.018 & 23.000 & 22.582 & \nodata \\ 
\enddata
\label{tab:table5}
\end{deluxetable}

\clearpage

\clearpage
\newpage
\begin{deluxetable}{ccccc}
\tablecaption{Ridge lines for the globular cluster M$\,$71.}
\tabletypesize{\scriptsize}
\tablewidth{0pt}
\tablehead{\colhead{$\uprime$}  &  
           \colhead{$\gprime$}  &  
           \colhead{$\rprime$}  &  
           \colhead{$\iprime$}  &  
           \colhead{$\zprime$}  }  
\startdata
 17.260 & 13.047 & 11.500 & 10.646 & 10.146 \\ 
 16.920 & 13.344 & 12.000 & 11.360 & 10.973 \\ 
 16.776 & 13.703 & 12.500 & 11.934 & 11.607 \\ 
 16.863 & 14.102 & 13.000 & 12.489 & 12.201 \\ 
 17.046 & 14.519 & 13.500 & 13.036 & 12.775 \\ 
 17.283 & 14.953 & 14.000 & 13.561 & 13.321 \\ 
 17.561 & 15.407 & 14.500 & 14.077 & 13.853 \\ 
 17.877 & 15.871 & 15.000 & 14.588 & 14.377 \\ 
 18.233 & 16.337 & 15.500 & 15.095 & 14.893 \\ 
 18.633 & 16.813 & 16.000 & 15.601 & 15.405 \\ 
 19.062 & 17.295 & 16.500 & 16.106 & 15.913 \\ 
 19.232 & 17.489 & 16.700 & 16.309 & 16.116 \\ 
 19.403 & 17.681 & 16.900 & 16.519 & 16.329 \\ 
 19.518 & 17.856 & 17.100 & 16.763 & 16.587 \\ 
 19.397 & 17.886 & 17.200 & 16.894 & 16.743 \\ 
 19.285 & 17.932 & 17.300 & 17.010 & 16.876 \\ 
 19.294 & 18.011 & 17.400 & 17.117 & 16.993 \\ 
 19.448 & 18.193 & 17.600 & 17.320 & 17.204 \\ 
 19.640 & 18.389 & 17.800 & 17.520 & 17.405 \\ 
 19.855 & 18.594 & 18.000 & 17.717 & 17.601 \\ 
 20.086 & 18.806 & 18.200 & 17.912 & 17.793 \\ 
 20.337 & 19.022 & 18.400 & 18.106 & 17.981 \\ 
 20.609 & 19.245 & 18.600 & 18.297 & 18.165 \\ 
 20.893 & 19.470 & 18.800 & 18.487 & 18.347 \\ 
 21.193 & 19.697 & 19.000 & 18.674 & 18.525 \\ 
 22.029 & 20.287 & 19.500 & 19.132 & 18.959 \\ 
 22.977 & 20.908 & 20.000 & 19.576 & 19.370 \\ 
 24.021 & 21.571 & 20.500 & 20.001 & 19.753 \\ 
 25.075 & 22.245 & 21.000 & 20.421 & 20.123 \\ 
 \nodata & 22.883 & 21.500 & 20.835 & 20.477 \\ 
 \nodata & 23.480 & 22.000 & 21.240 & 20.826 \\ 
 \nodata & 24.009 & 22.500 & 21.632 & 21.188 \\ 
 \nodata & 24.509 & 23.000 & 22.132 & 21.685 \\ 
\enddata
\label{tab:table6}
\end{deluxetable}

\clearpage

\clearpage
\newpage
\begin{deluxetable}{ccccc}
\tablecaption{Ridge lines for the open cluster NGC$\,$6791.}
\tabletypesize{\scriptsize}
\tablewidth{0pt}
\tablehead{\colhead{$\uprime$}  &  
           \colhead{$\gprime$}  &  
           \colhead{$\rprime$}  &  
           \colhead{$\iprime$}  &  
           \colhead{$\zprime$}  }  
\startdata
 \nodata & 14.514 & 13.100 & 12.370 & 11.906 \\ 
 \nodata & 14.898 & 13.600 & 13.057 & 12.690 \\ 
 \nodata & 15.314 & 14.100 & 13.632 & 13.325 \\ 
 \nodata & 15.745 & 14.600 & 14.176 & 13.915 \\ 
 \nodata & 16.186 & 15.100 & 14.711 & 14.481 \\ 
 \nodata & 16.640 & 15.600 & 15.237 & 15.029 \\ 
 \nodata & 17.106 & 16.100 & 15.752 & 15.556 \\ 
 \nodata & 17.577 & 16.600 & 16.259 & 16.072 \\ 
 \nodata & 17.765 & 16.800 & 16.461 & 16.276 \\ 
 \nodata & 17.946 & 17.000 & 16.665 & 16.482 \\ 
 \nodata & 18.032 & 17.120 & 16.801 & 16.626 \\ 
 \nodata & 17.966 & 17.120 & 16.833 & 16.687 \\ 
 \nodata & 17.842 & 17.060 & 16.803 & 16.679 \\ 
 \nodata & 17.771 & 17.050 & 16.814 & 16.709 \\ 
 \nodata & 17.885 & 17.200 & 16.979 & 16.885 \\ 
 \nodata & 18.076 & 17.400 & 17.182 & 17.089 \\ 
 \nodata & 18.281 & 17.600 & 17.382 & 17.289 \\ 
 \nodata & 18.495 & 17.800 & 17.579 & 17.483 \\ 
 \nodata & 18.715 & 18.000 & 17.773 & 17.673 \\ 
 \nodata & 18.940 & 18.200 & 17.965 & 17.858 \\ 
 \nodata & 19.166 & 18.400 & 18.154 & 18.040 \\ 
 \nodata & 19.402 & 18.600 & 18.339 & 18.214 \\ 
 \nodata & 20.014 & 19.100 & 18.794 & 18.638 \\ 
 \nodata & 20.658 & 19.600 & 19.227 & 19.031 \\ 
 \nodata & 21.311 & 20.100 & 19.629 & 19.387 \\ 
 \nodata & 21.942 & 20.600 & 20.008 & 19.707 \\ 
 \nodata & 22.525 & 21.100 & 20.365 & 19.996 \\ 
 \nodata & 23.048 & 21.600 & 20.714 & 20.276 \\ 
 \nodata & 23.546 & 22.100 & 21.057 & 20.541 \\ 
 \nodata & 24.033 & 22.600 & 21.400 & 20.806 \\ 
 \nodata & 24.522 & 23.100 & 21.768 & 21.108 \\ 
\enddata
\label{tab:table7}
\end{deluxetable}

\clearpage

\clearpage

\clearpage

\clearpage

\clearpage

\clearpage

\clearpage

\clearpage

\clearpage

\clearpage

\clearpage

\clearpage

\clearpage

\clearpage

\end{document}